\documentclass[11pt]{article}
\usepackage[utf8]{inputenc}
\usepackage[T1]{fontenc}

\DeclareMathAlphabet{\mathpzc}{OT1}{pzc}{m}{it}

\usepackage{amsmath,amssymb}
\usepackage[active]{srcltx}

\usepackage[all]{xy}
\input diagxy

\usepackage{tikz-cd}
\usetikzlibrary{cd}

\usepackage{xcolor}
\usepackage{enumerate}
\usepackage{dsfont}
\usepackage{mathrsfs}
\usepackage{comment}
\usepackage{array}

\font\fr=eufm10 scaled \magstep 1 
\def\beq{\begin{equation}}
\def\eeq{\end{equation}}
\def\bea{\begin{eqnarray}}
\def\eea{\end{eqnarray}}
\def\beann{\begin{eqnarray*}}
\def\eeann{\end{eqnarray*}}
\def\beasn{\begin{sneqnarray}}
\def\eeasn{\end{sneqnarray}}
\def\ben{\begin{enumerate}}
\def\een{\end{enumerate}}
\def\bit{\begin{itemize}}
\def\eit{\end{itemize}}

\def\derpar#1#2{\frac{\partial{#1}}{\partial{#2}}}

\def\moment#1#2#3{{#1}_{#2}, \ldots, {#1}_{#3}}
\def\qed{\ifvmode\Realemovelastskip\fi
{\unskip\nobreak\hfil\penalty50\hbox{}\nobreak\hfil \hbox{\vrule
height1.2ex width1.2ex}\parfillskip=0pt \finalhyphendemerits=0
\par\smallskip}}

\def\vf{\text{\fr X}}
\def\df{{\mit\Omega}}
\def\Lag{{\cal L}}
\def\L{{\cal L}}

\def\d{{\rm d}}

\def\Tan{{\rm T}}
\def\Lie{\mathop{\rm L}\nolimits}
\def\inn{\mathop{i}\nolimits}

\makeatletter
\DeclareOldFontCommand{\rm}{\normalfont\rmfamily}{\mathrm}
\DeclareOldFontCommand{\sf}{\normalfont\sffamily}{\mathsf}
\DeclareOldFontCommand{\tt}{\normalfont\ttfamily}{\mathtt}
\DeclareOldFontCommand{\bf}{\normalfont\bfseries}{\mathbf}
\DeclareOldFontCommand{\it}{\normalfont\itshape}{\mathit}
\DeclareOldFontCommand{\sl}{\normalfont\slshape}{\@nomath\sl}
\DeclareOldFontCommand{\sc}{\normalfont\scshape}{\@nomath\sc}
\makeatother

\usepackage[english]{babel}
\usepackage[utf8]{inputenc}
\usepackage[T1]{fontenc}
\usepackage{eucal}
\usepackage{microtype}
\usepackage{csquotes}
\usepackage{amsthm}
\usepackage{mathtools}
\mathtoolsset{showonlyrefs}
\usepackage{amssymb,amsfonts,stmaryrd}
\usepackage{tikz}
\usetikzlibrary{cd}
\usepackage{xparse}

\usepackage{todo}

\RequirePackage{hyperref}

\theoremstyle{plain}
\newtheorem{theorem}{Theorem}[section]
\newtheorem*{theorem*}{Theorem}

\newtheorem*{lemma*}{Lemma}
\newtheorem{proposition}[theorem]{Proposition}
\newtheorem*{proposition*}{Proposition}

\newtheorem*{corollary*}{Corollary}
\newtheorem{definition}[theorem]{Definition}
\newtheorem*{definition*}{Definition}

\newtheorem*{example*}{Example}
\newtheorem{remark}[theorem]{Remark}
\newtheorem*{remark*}{Remark}

\newtheorem*{conjecture*}{Conjecture}

\newtheorem*{problem*}{Problem}

\newcommand*{\R}{\mathbb{R}}

\renewcommand{\L}{\mathcal{L}}

\renewcommand{\d}{\mathrm{d}}

\title{\vskip -10mm
\sc Canonical Lifts in Multisymplectic De Donder-Weyl Hamiltonian Field Theories}
\author{\sffamily 
\sc 
$^a$ Arnoldo Guerra IV
\thanks{aguerrgu54@alumnes.ub.edu\,({\it ORCID}:\,0000-0001-8738-274X).} ,
$^b$ Narciso Rom\'an-Roy
\thanks{narciso.roman@upc.edu\,({\it ORCID}:\,0000-0003-3663-9861).} .
\\[1ex]
\normalsize\itshape\sffamily 
$^a$Departament de F\'isica Qu\`antica i Astrof\'isica,
Universitat de Barcelona, 
Barcelona, Spain.
\\[1ex]
\normalsize\itshape\sffamily 
$^b$Department of Mathematics,
Universitat Polit\`ecnica de Catalunya,
Barcelona, Spain.
}

\begin{document}

\maketitle

\begin{abstract}
We define canonical lifts of vector fields to the multisymplectic multimomentum bundles of De Donder--Weyl Hamiltonian (first-order) field theories and to the appropriate premultisymplectic embedded constraint submanifolds on which singular field theories are studied. These new canonical lifts are used to study the so-called {\sl natural Noether symmetries} present in both regular and singular Hamiltonian field theories along with their associated conserved quantities obtained from Noether's theorem. The {\sl Klein--Gordon field},
the {\sl Polyakov bosonic string}, and {\sl Einstein--Cartan gravity in $3+1$ dimensions} are analyzed in depth as applications of these concepts; as a peripheral result obtained in the analysis of the bosonic string, we provide a new geometrical interpretation of the well-known {\sl Virasoro constraint}. 
\end{abstract}


\noindent\textbf{MSC\,2020 codes:}
{\it Primary}: 53D42, 70S05, 83C05; {\it Secondary}: 35Q75, 35Q76, 53Z05, 70H50,  83C99.


\pagestyle{myheadings}
\markright{\small\itshape\sffamily 
{\rm A. Guerra IV, N. Rom\'an-Roy}:
Canonical Lifts in Multisymplectic De Donder--Weyl Field Theories.}


 \bigskip
 
{\setcounter{tocdepth}{2}
\def\baselinestretch{1}
\small
\def\addvspace#1{\vskip 1pt}
\parskip 0pt plus 0.1mm
\tableofcontents
}

\newpage

\section{Introduction}

Over the years, the most important symmetries exhibited by classical field theories have arguably been {\sl Noether symmetries}. 
The vast majority of Noether symmetries in physical field theories are of a specific kind, often referred to in the literature as {\sl natural symmetries}, generated by Lie group actions which are {\sl canonical lifts} from a configuration manifold of the field theory under investigation to some phase space on which the field theory is to be studied. 
We wish to discuss such symmetries in this work in a new geometric context, relevant to the De Donder--Weyl Hamiltonian treatment of field theories \cite{dD-1930,We-1935}.

The De Donder--Weyl formulation of classical field theories has been proven to be equivalent to the well-known Lagrangian formulation. 
In fact, one of the main characteristics of the De Donder--Weyl formalism is that the equations of motion are given as two sets of first-order partial differential equations, known as the Hamilton--De Donder--Weyl equations, which are equivalent to the Euler--Lagrange equations while maintaining manifest Lorentz covariance for Lorentz invariant field theories.
The relevant Legendre map arises from constructing the {\sl canonical multimomenta} which are obtained from the variation of the Lagrangian function with respect to spacetime derivatives of the fields (sometimes called {\sl multivelocities}),
as opposed to varying with respect to only time derivatives of the fields.
However, the structure of field theories that are singular under this Legendre transform remained unknown for many years. 

In more recent years, the Lagrangian and De Donder--Weyl Hamiltonian formulations of field theories have been posed in more geometrically formal terms on {\sl jet} and {\sl affine dual jet} bundles respectively \cite{art:Aldaya_Azcarraga78_2,AA-80,CCI-91,LMM-96,EMR-99b,EMR-96,EMR-00b,Gc-73,GS-73,HK-04,art:Roman09,book:Saunders89}.
This formulation begins from constructing a finite dimensional configuration bundle, usually, over space-time in which the fields of the theory are given by local sections of the bundle structure.
Then, the jet bundle on which the Lagrangian formalism takes place is constructed from the projection map of the configuration bundle and is also a fiber bundle over spacetime. 
Furthermore, the so-called affine dual jet bundle, which is
also commonly referred to as the
{\sl restricted multimomentum bundle}, is where the equivalent De Donder--Weyl formulation takes place (and is also a fiber bundle over spacetime).
It should be noted that in the case of the Polyakov string, the base space of the entire aforementioned fiber bundle structure is instead the string worldsheet.
An important advantage of working on the jet and multimomentum bundles is that they are finite dimensional manifolds. 
In the physics literature, the Lagrangian formulation of field theories usually takes place on the space of sections of the relevant jet bundle over spacetime, giving an infinite dimensional phase space on which the variational calculus is carried out. 
Similarly, the original construction of the De Donder--Weyl formalism takes place on an infinite dimensional space of sections of the relevant multimomentum bundle.
The variational calculus is then carried out by taking functional variations of an {\sl action functional} that is defined on these infinite dimensional spaces.

However, the variational calculus can instead be carried out on the jet and multimomentum bundles themselves, referred to in this work as the {\sl Lagrangian} and {\sl Hamiltonian multiphase spaces} respectively. 
The geometric method for carrying out the variational calculus on the multiphase spaces involves the use of the multisymplectic or premultisymplectic structures 
which these multiphase spaces are endowed with
\cite{art:Aldaya_Azcarraga78_2,EMR-00b,Gc-73,GS-73}. 
The Hamiltonian multiphase space 
inherits its (pre)multisymplectic structure
from the {\sl extended multimomentum bundle}, a bundle of differential forms that comes equipped with a canonical exact multisymplectic structure and is a fiber bundle over the restricted multimomentum bundle on which the Hamiltonian variational calculus and geometric constraint analysis is carried out.
On the other hand, the (pre)multisymplectic form on the Lagrangian multiphase space is constructed from the Lagrangian
and is related to the canonical form on the extended multimomentum bundle by an {\sl extended Legendre transform}.
Then,
the degeneracy of the (pre)multisymplectic forms on the Lagrangian and Hamiltonian multiphase spaces depends on the regularity of the theory under investigation, which is specified by a Lagrangian or a De Donder--Weyl Hamiltonian function. 
Moreover, for singular field theories, the premultisymplectic analysis on multiphase spaces provides us with a method to perform a geometric constraint analysis.
This constraint analysis was introduced in \cite{LMM-96,LMMMR-2005} and was detailed further and applied to examples of field theories in \cite{GR1,GR2,GGR-2022}.
This geometric formalism can be viewed as a multivariable generalization of the presymplectic formulation of time-dependent mechanics; 
(see also \cite{Kanatchikov1} for an alternate proposed geometric formalism).

Additionally, the notion of canonical lifts of Lie group actions and their generating vector fields on jet manifolds
has been well understood for many years (see e.g. \cite{book:Saunders89}).
Furthermore, symmetries (and more specifically, natural symmetries) on jet bundles have been well understood in terms of (pre)multisymplectic geometry \cite{art:Aldaya_Azcarraga78_2,AA-78,art:deLeon_etal2004,De-77,EMR-96,EMR-99b,FS-2012,Gc-73,GS-73,GIMMSY,Krupka,RWZ-2016,book:Saunders89};
for an overview of symmetries in the (pre)multisymplectic Lagrangian setting, see \cite{art:GPR-2016,GR-3,GR-2023}.
However, to our knowledge, the notions of canonical lifts and natural symmetries on the multimomentum bundles 
of (pre)multisymplectic De Donder--Weyl Hamiltonian field theories have not been rigorously studied in previous literature.

In this work, we present a formal mathematical definition of the canonical lift of a configuration bundle diffeomorphism
and of a vector field 
on the configuration bundle of a field theory to the extended and restricted multimomentum bundles on which the multisymplectic De Donder--Weyl Hamiltonian formulation of field theories takes place. 
The notion of natural symmetries in the multisymplectic De Donder--Weyl framework thereby follows, along with a variety of properties and new theorems. 
We also give a preliminary definition of canonical lifts for field theories that are singular (or, more specifically, {\sl almost-regular} \cite{LMM-96,Sa-95}) in the De Donder--Weyl sense. 
The De Donder--Weyl Hamiltonian formulation of almost-regular field theories takes place on an embedded primary constraint submanifold of the restricted multimomentum bundle.
However, in order to arrive at the definition we give 
for a canonical lift to embedded submanifolds of a group action depends on the invariance of the primary constraint submanifold under the lifted group action on the restricted multimomentum bundle.
Nevertheless, this requirement could be very restrictive as the last example
analyzed in Section \ref{Einstein-Cartan} shows.

Prior to this work, the usual way to obtain the appropriate symmetries resulting from canonical lifts in the De Donder--Weyl setting was to first work out the full analysis in the Lagrangian setting, including the variational calculus and constraint analysis, and then project the constraint and symmetry structures onto the Hamiltonian setting via the Legendre transform \cite{GR-2023}. 
The results presented in this paper allow us to go directly to the De Donder--Weyl formalism for classical field theories without having to first walk through the entire Lagrangian formulation.

This paper is organized as follows:
First, in Section 2, a review of the (pre)multisymplectic De Donder--Weyl Hamiltonian formalism for field theories is given; we also give a brief review on canonical lifts of vector fields from spacetime to the configuration bundle of some generic field theory to be studied. 
The main results of this work are in Sections 3 and 4.
Section 3 is devoted to defining and characterizing the canonical lifts of vector fields from the configuration bundle to the {\sl extended} and {\sl restricted multimomentum bundles} of a De Donder--Weyl Hamiltonian field theory. 
We also provide the definition and characterization of the analogous canonical lifts to the
embedded constraint submanifolds of the extended and restricted multimentum bundles on which singular theories are studied \cite{LMM-96,LMMMR-2005}.
Next, in Section 4 we study symmetries for regular and singular Hamiltonian field theories,
paying special attention to those symmetries generated by vector fields
that are canonical lifts (i.e. natural symmetries);
the statement of Noether´s theorem, involving conserved quantities (multimomentum maps) is given explicitly.  
Finally, in Section 5 we study some relevant field theories in theoretical physics to illustrate the new mathematical concepts introduced in this work.
The first example theory is the {\sl Klein--Gordon field}, which is regular.
Next, we study the {\sl Polyakov bosonic string} and {\sl Einstein–Cartan gravity in $3+1$ dimensions}, both of which are singular.
The full premultisymplectic constraint analysis is worked out for both singular theories. 
Furthermore, we give the appropriate construction of canonical lifts, and the resulting natural symmetries, in both the Lagrangian and Hamiltonian formalisms for all field theories discussed in this paper.

In the case of the bosonic string, we find from the premultisymplectic constraint analysis that the 
Virasoro constraint arises as a {\sl compatibility constraint}.
The specific geometric relationship between our premultisymplectic constraint analysis and the more well-known canonical Dirac Hamiltonian constraint analysis (which is given elegantly for the Polyakov string in \cite{Gomis-String}) is still unknown and is left for future work.

All the manifolds are real, second countable, and of class $ C^\infty$, and the mappings are assumed to be smooth.
Sum over crossed repeated indices is understood.
Throughout the paper, spacetime $M$ has local coordinates denoted $x^\mu$ and Minkowski metric $\eta_{\mu\nu}$ with signature $(-+...+)$; the spacetime partial derivatives are denoted as $\partial_\mu = \partial/\partial x^\mu$.

\section{Preliminaries}

\subsection{Multisymplectic Hamiltonian Field Theory}
\label{MHFT}

(See, for instance, \cite{CCI-91,LMM-96,EMR-99b,
GIMMSY,HK-04,art:Roman09} for more details).

Recall that,
given a differentiable manifold $\mathcal{M}$,
 a differential form $\Omega\in\df^m(\mathcal{M})$ ($m\geq 2$) is
{\bf $1$-nondegenerate} if, for every ${\rm p}\in\mathcal{M}$ and $Y\in\vf(\mathcal{M})$,
it follows that $\inn(Y)\Omega\vert_{\rm p}=0\ \Longleftrightarrow\ Y\vert_{\rm p}=0$.
If the differential form $\Omega\in\df^m(\mathcal{M})$ is closed and $1$-nondegenerate, then it is called a {\sl\textbf{multisymplectic form}} 
and $(\mathcal{M},\Omega)$ is called a {\sl\textbf{multisymplectic manifold}};
furthermore, when $\Omega$ is closed and $1$-degenerate, it is called a {\sl\textbf{premultisymplectic form}}
and $(\mathcal{M},\Omega)$ is called a {\sl\textbf{premultisymplectic manifold}}.

\begin{paragraph}{Regular Hamiltonian field theories:}

Let the {\sl configuration bundle} of a first-order field theory
be a fiber bundle $\pi\colon E\to M$, 
where $M$ is an orientable manifold ($\dim\,M=m>1$, $\dim\,E=n+m$)
with volume form $\omega\in\df^m(M)$.
The manifold $M$ is usually spacetime which also comes with a pseudo-Riemannian metric; in the case of string theory, the base space is the string worldsheet which comes with an induced metric.
Local coordinates adapted to the bundle structure are denoted  $(x^\mu, y^A)$ with $0\leq\mu\leq m-1$, $1\leq A\leq n$;
the volume form is expressed in these coordinates as $\omega=\d x^0\wedge\ldots\wedge\d x^{m-1}\equiv\d^mx$.
The fields $y^A(x)$ of the theory under investigation are given by the local sections 
on the configuration bundle, $\phi:M\rightarrow E:x^\mu\mapsto (x^\mu,y^A(x))$.

The geometric framework for the {\sl De Donder--Weyl Hamiltonian formulation} of first-order field theories
is constructed by first introducing the {\sl \textbf{extended multimomentum bundle}}
${\cal M}\pi\equiv\Lambda_2^m\Tan^*E$
which is the bundle of $m$-forms on
$E$ vanishing by the action of two $\pi$-vertical vector fields.
This manifold has natural coordinates $(x^\mu,\,y^A,p_A^\mu, p)$, 
and hence $\text{dim}\,{\cal M}\pi=m+n+mn+1$.
As ${\cal M}\pi\equiv\Lambda_2^m\Tan^*E$ is a subbundle of $\Lambda^m\Tan^*E$, the {\sl multicotangent bundle} of $E$ of order $m$,
it follows that ${\cal M}\pi$ is endowed with a canonical form,
the ``tautological $m$-form'' $\Theta\in\df^m({\cal M}\pi)$,
which is defined as follows:
if ${\rm p}\equiv(y,\zeta)\in{\cal M}\pi$ with $y\in E$ 
and
$\zeta\in\Lambda_2^m\Tan_y^*E$, then,
for every 
$X_0,\ldots,X_{m-1}\in\Tan_{(y,\zeta)}({\cal M}\pi)$,
\beq
\label{Thetadef}
\Theta_{(y,\zeta)}(\moment{X}{0}{m-1}):=
\zeta(\Tan_{(y,\zeta)}\widetilde\sigma(X_0),\dotsc ,\Tan_{(y,\zeta)}\widetilde\sigma(X_{m-1})) \ ,
\eeq
where $\widetilde\sigma\colon{\cal M}\pi\to E$ is the natural projection.
Consequently, $\Omega=-\d\Theta\in\df^{m+1}({\cal M}\pi)$ is a multisymplectic form.
The local expressions of these forms are,
\begin{equation}
\label{coordforms}
  \Theta=p_A^\mu\d y^A\wedge\d^{m-1}x_\mu+p\,\d^mx
  \quad , \quad
  \Omega = -\d p_A^\mu\wedge\d y^A\wedge\d^{m-1}x_\mu-\d p\wedge\d^mx \ ,
\end{equation}
and are known as the {\sl \textbf{multimomentum Liouville 
$m$ and $(m+1)$-forms}} of ${\cal M}\pi$ respectively.

The quotient bundle $J^1\pi^*:={\cal M}\pi/\Lambda^m_1(\Tan^*{\rm E})$,
where $\Lambda_1^m(\Tan^*{\rm E})$ denotes the bundle of $\pi$-semibasic $m$-forms on ${\rm E}$,
is called the {\sl\textbf{(restricted) multimomentum bundle}} of $E$
and has natural coordinates $(x^\mu,y^A,p^\mu_A)$.
Sections ${\rm h}\colon J^1\pi^*\to{\cal M}\pi$ 
of the bundle $\sigma:{\cal M}\pi\rightarrow J^1\pi^*$
are called {\sl \textbf{Hamiltonian sections}}.
Hamiltonian sections can be used to define the forms,
\begin{equation}
\label{defThetaOmegaH}
 \Theta_{\mathscr H}:={\rm h}^*\Theta\in\df^m(J^1\pi^*)\ ,\qquad
 \Omega_{\mathscr H}:=-\d\Theta_{\mathscr H}={\rm h}^*\Omega\in\df^{m+1}(J^1\pi^*) \ ,
\end{equation}
which are called the {\sl\textbf{Hamilton--Cartan $m$ and $(m+1)$ forms}} of $J^1\pi^*$
associated with the Hamiltonian section ${\rm h}$.
A Hamiltonian section is locally specified by
${\rm h}(x^\nu,y^A,p^\nu_A)=(x^\nu,y^A,p^\nu_A,p=-{\mathscr H}(x^\gamma,y^B,p_B^\gamma))$,
where $\mathscr{H}$ is a {\sl (local) Hamiltonian function},
sometimes known as a {\sl\textbf{De Donder--Weyl Hamiltonian function}}. 
Then, the local expressions for the Hamilton--Cartan forms are:
\begin{equation*}
 \Theta_\mathscr{H} = p_A^\mu\d y^A\wedge\d^{m-1}x_\mu -\mathscr{ H}\d^mx
\quad  , \quad
 \Omega_\mathscr{H} = -\d p_A^\mu\wedge\d y^A\wedge\d^{m-1}x_\mu +
 \d \mathscr{H}\wedge\d^mx \ .
\end{equation*}
The form $\Omega_\mathscr{H}$ is multisymplectic
and the pair $(J^1\pi^*,\Omega_{\mathscr{H}})$
is called a {\sl \textbf{regular Hamiltonian system}}.

The natural projections (submersions) are listed as follows,
\beann
\tau:J^1\pi^*\rightarrow E&:&(x^\mu,y^A,p^\mu_A)\mapsto (x^\mu,y^A) \ ,
\\
\bar{\tau}:J^1\pi^*\rightarrow M&:&(x^\mu,y^A,p^\mu_A)\mapsto (x^\mu) \ ,
\\
\sigma:{\cal M}\pi\rightarrow J^1\pi^*&:&(x^\mu,y^A,p^\mu_A,p)\mapsto (x^\mu,y^A,p^\mu_A) \ ,
\\
\widetilde\sigma=\tau\circ\sigma\colon{\cal M}\pi\to E &: & (x^\mu,y^A,p^\mu_A,p)\mapsto (x^\mu,y^A) \ ,
\\
\overline{\sigma}=\pi\circ\widetilde\sigma\colon{\cal M}\pi\to M &: & (x^\mu,y^A,p^\mu_A,p)\mapsto (x^\mu) \ ,
\eeann
and are depicted in the following diagram:
\begin{equation}\label{diagram1}
    \begin{tikzcd}[column sep=1.75cm, row sep=1.25cm]
       \mathcal{M}\pi \arrow[rr, "\sigma"] \arrow[dr, "\widetilde\sigma"] \arrow[rdd, swap, "\overline{\sigma}"] & & J^1\pi^\ast \arrow[dl, swap, "\tau"] \arrow[ldd, "\overline\tau"] \\
       & E \arrow[d, "\pi"] \\
       & M
    \end{tikzcd}
\end{equation}

The field equations are obtained from a variational principle on $J^1\pi^*$ in which
the {\sl action} is a functional on the set of sections $\Gamma(\overline{\tau})$ given by
$\displaystyle 
\Gamma(M,J^1\pi^*)\rightarrow \R:\psi \mapsto \int_{M}\psi^*\Theta_\mathscr{H}$.
The objective is to find sections that are critical (stationary),
\begin{equation*}
\left.\frac{\text{d}}{\text{d}s}\right\vert_{s=0}\int_{M} \psi^*_s\Theta_{\mathscr{H}}=0 \ ,
\end{equation*}
for variations $\psi_s=\eta_s\circ \psi$ of $\psi$, 
where $\eta_s$ is the flow of a $\overline{\tau}$-vertical vector field on $J^1\pi^*$ compactly supported on $M$.
Critical sections are characterized in the following equivalent ways:
\ben
\item
$\psi^*i(X)\Omega_{\mathscr{H}}=0$,
for every $X\in \mathfrak{X}(J^1\pi^*)$.
See also \cite{EMR-00b,art:Roman09} for other equivalent statements.
\item
$\psi$ is an integral section of a class of integrable,
locally decomposable, $\bar{\tau}$-transverse multivector fields 
$\{{\bf X}_{\mathscr{H}}\} \subset \mathfrak{X}^m(J^1\pi^*)$ 
(see the appendix \ref{append}) which satisfy 
\begin{equation}
\label{equationHEOM}
\inn({\bf X}_{\mathscr{H}})\Omega_{\mathscr{H}}=0
\quad , \quad \mbox{\rm  for every ${\bf X}_{\mathscr{H}} \in  \{{\bf X}_{\mathscr{H}}\}$} \ ;
\end{equation}
then, if $\psi$ is an integral section of ${\bf X}_{\mathscr{H}}$,
the relation $\psi^{(m)}={\bf X}_{\mathscr{H}}\circ\psi$ holds
(see Appendix \ref{append}) 
and the corresponding equation for these integral sections is
$$
\inn(\psi^{(m)})(\Omega_{\mathscr{H}}\circ\psi)=0\ .
$$
Furthermore, it is possible to choose a representative of the class $\{{\bf X}_{\mathscr{H}}\}$
which is a normalized, $\overline{\tau}$-transverse, multivector field such that 
\begin{equation}
\label{equationHEOM1}
i({\bf X}_{\mathscr{H}})\omega=1 \ .
\end{equation}
\item
Using a natural system of coordinates on $J^1\pi^*$, 
the section $\psi$ satisfies the {\sl \textbf{Hamilton--De Donder--Weyl equations}} expressed as:
$$
\frac{\partial (y^A\circ \psi)}{\partial x^\mu} = \frac{\partial \mathscr{H}}{\partial p_A^\mu} \circ \psi
\quad , \quad \frac{\partial (p_A^\mu \circ \psi)}{\partial x^\mu}=-\frac{\partial \mathscr{H}}{\partial y^A} \circ \psi \ .
$$
\een
\end{paragraph}

\begin{paragraph}{Regular Lagrangian field theories:}
\label{canreg}

The Lagrangian formulation of a regular first-order field theory takes place
on the first-order jet bundle $J^1\pi$ of $\pi\colon E\to M$,
which is called the {\sl (first-order) multivelocity phase space}
and has local coordinates denoted $\left(x^\mu,y^A,y_\mu^A\right)$;
see \cite{art:Aldaya_Azcarraga78_2,EMR-96,art:Echeverria_Munoz_Roman98,Gc-73,art:GPR-2016,GS-73,book:Saunders89} for details.
A Lagrangian density ${\bf L}\in\df^{m}(J^1\pi)$ is an $m$-form 
such that ${\bf L}=\mathscr{L}\,\omega$,
where $\mathscr{L}\in C^\infty(J^1\pi)$ is called the {\sl Lagrangian function}
and $\omega$ denotes the pull-back of the volume form from $M$ to $J^1\pi$.
Using the canonical endomorphism which $J^1\pi$ is endowed with,
define the {\sl \textbf{Poincar\'e--Cartan $m$}} and {\sl \textbf{$(m+1)$-forms}}
$\Theta_\mathscr{L}\in\df^{m}(J^1\pi)$ and
$\Omega_\mathscr{L}=-\d\Theta_\mathscr{L}\in\df^{m+1}(J^1\pi)$,
with coordinate expression:
\begin{align}
\Theta_{\mathscr{L}} &= \frac{\partial \mathscr{L}}{\partial y^A_\mu}\d y^A\wedge\d^{m-1}x_\mu -\left(\frac{\partial \mathscr{L}}{\partial y^A_\mu}y^A_\mu-\mathscr{L}\right)\d^m x \equiv \frac{\partial \mathscr{L}}{\partial y^A_\mu}\d y^A\wedge\d^{m-1}x_\mu -E_\mathscr{L}\,\d^m x \ , \\
\Omega_\mathscr{L} &= - \text{d}\Theta_\mathscr{L}= -\text{d}\left(\frac{\partial \mathscr{L}}{\partial y^A_\mu}\right)\wedge\d y^A\wedge\d^{m-1}x_\mu +\text{d}E_\mathscr{L}\,\d^m x \ . 
\end{align}
The pair $(J^1\pi,\Omega_\mathscr{L})$ is called a {\sl \textbf{first-order Lagrangian system}}
which is said to be {\bf regular} if $\Omega_\mathscr{L}$ is a multisymplectic form and is {\sl singular} otherwise
(that is, when $\Omega_\mathscr{L}$ is premultisymplectic).
The regularity condition is locally equivalent to demanding that
the generalized Hessian matrix
$\displaystyle\frac{\partial^2\mathscr{L}}
{\partial y^A_\mu\partial y^B_\nu}$
be non-singular everywhere on $J^1\pi$.
The {\sl \textbf{extended Legendre map}}, $\widetilde{\mathscr{FL}}\colon J^1\pi\to {\cal M}\pi$,
and the {\sl \textbf{Legendre map}}, $\mathscr{FL} :=\sigma\circ\widetilde{\mathscr{FL}}\colon J^1\pi\to J^1\pi^*$,
associated with a Lagrangian $\mathscr{L}\in C^\infty(J^1\pi)$
are given locally as
 $$
 \begin{array}{ccccccc}
\widetilde{\mathscr{FL}}^*x^\nu = x^\nu &, &
  \widetilde{{\mathscr{FL}}}^*y^A = y^A &\ , \ &
\widetilde{\mathscr{FL}}^*p_A^\nu =\displaystyle\frac{\partial\mathscr{L}}{\partial y^A_\nu}
 & , \ &
 \widetilde{\mathscr{FL}}^*p =\mathscr{L}-\displaystyle y^A_\nu\frac{\partial\mathscr{L}}{\partial y^A_\nu}
 \\
 \mathscr{FL}^*x^\nu = x^\nu &\ , \ &
 \mathscr{FL}^*y^A = y^A &\ , \ &
 \mathscr{FL}^*p_A^\nu =\displaystyle\frac{\partial\mathscr{L}}{\partial y^A_\nu}\ . & &
 \end{array}
 $$
The Lagrangian function is regular
if, and only if, $\mathscr{FL}$ is a local diffeomorphism.
As a particular case, $\mathscr{L}$ is {\sl \textbf{hyper-regular}}
if $\mathscr{FL}$ is a global diffeomorphism.
It is worth noting that the regularity of a field theory is a quite demanding condition that is not met by many of the important physical field theories, e.g. Yang-Mills and General Relativity among others.

In the (hyper-)regular case, the map
${\rm h}:=\widetilde{\mathscr{FL}}\circ\mathscr{FL}^{-1}$ is a
Hamiltonian section which is associated with $\mathscr{L}$
and, as $\displaystyle E_\mathscr{L}=\frac{\partial \mathscr{L}}{\partial y^A_\mu}y^A_\mu-\mathscr{L}$,
the De Donder--Weyl Hamiltonian function is given as,
$$
\mathscr{H}=(\mathscr{FL}^{-1})^*E_\mathscr{L}=p^\mu_A
(\mathscr{FL}^{-1})^*y^A_\mu-(\mathscr{FL}^{-1})^*\mathscr{L}\in C^\infty(J^1\pi^*)\ .
$$
The following diagram illustrates the relations between the multisymplectic spaces discussed thus far:
\begin{equation}\label{diagram2}
    \begin{tikzcd}[column sep=1.75cm, row sep=1.25cm]
        & \mathcal{M}\pi \arrow[d, swap, "\sigma"] \\
        J^1\pi \arrow[ur, "\widetilde{\mathscr{FL}}"] \arrow[r, swap, "\mathscr{FL}"] & J^1\pi^\ast \arrow[u, swap, bend right=30, pos=0.49, "\mathrm{h}"]
    \end{tikzcd}
\end{equation}
It follows that $\mathscr{FL}^*\Theta_\mathscr{H}=\Theta_\mathscr{L}$
and $\mathscr{FL}^*\Omega_\mathscr{H}=\Omega_\mathscr{L}$.

\noindent The pair $(J^1\pi^*,\Omega_\mathscr{H})$
 is the {\sl Hamiltonian system associated with the regular Lagrangian system} $(J^1\pi,\Omega_\mathscr{L})$.

The variational principle takes place on $J^1\pi$ and
the {\sl action} is a functional on the set of sections $\Gamma(\overline{\pi}^1)$ given by
$\displaystyle 
\Gamma(M,J^1\pi)\rightarrow \R:j^1\phi \mapsto \int_{M}(j^1\phi)^*\Theta_\mathscr{L}$.
The objective is to find sections that are critical (stationary),
\begin{equation*}
\left.\frac{\text{d}}{\text{d}s}\right\vert_{s=0}\int_{M} (j^1\phi)^*_s\Theta_{\mathscr{L}}=0 \ ,
\end{equation*}
for variations $j^1\phi_s=\eta_s\circ j^1\phi$ of $j^1\phi$,
where $\eta_s$ is the flow of a $\overline{\pi}^1$-vertical vector field on $E$ compactly supported on $M$.
Note that the sections $j^1\phi$ are {\sl holonomic} as they are canonical lifts of sections $\phi:M\to E$.
Critical sections are characterized in the following equivalent ways:
\ben
\item
$(j^1\phi)^*\inn(X)\Omega_{\mathscr{L}}=0$,
for every $X\in \mathfrak{X}(J^1\pi)$.
See also \cite{EMR-96,Gc-73,GS-73} for other equivalent statements.
\item
$j^1\phi$ is a holonomic integral section of a class of integrable,
locally decomposable, $\overline{\pi}^1$-transverse multivector fields 
$\{{\bf X}_{\mathscr{L}}\} \subset \mathfrak{X}^m(J^1\pi)$ 
(that is, a {\sl holonomic multivector field}) which satisfy 
\begin{equation}
\label{equationLEOM}
\inn({\bf X}_{\mathscr{L}})\Omega_{\mathscr{L}}=0
\quad , \quad \mbox{\rm  for every ${\bf X}_{\mathscr{L}} \in  \{{\bf X}_{\mathscr{L}}\}$} \ ,
\end{equation}
and the corresponding equation for the holonomic integral sections of ${\bf X}_{\mathscr{L}}$ is
$$
\inn\Big((j^1\phi)^{(m)}\Big)(\Omega_{\mathscr{L}}\circ j^1\phi)=0\ .
$$
Furthermore, it is possible to choose a representative of the class $\{{\bf X}_{\mathscr{L}}\}$
which is a normalized and $\overline{\tau}$-transverse multivector field so that 
\begin{equation}
\label{equationHEOM2}
\inn({\bf X}_{\mathscr{L}})\omega=1 \ .
\end{equation}
The local expression for ${\bf X}_{\mathscr{L}}$ is given as,

\begin{equation}
\label{XL}
{\bf X}_\mathscr{L}= \bigwedge^{m-1}_{\mu=0} X_\mu = \bigwedge^{m-1}_{\mu=0}  \left( \frac{\partial}{\partial x^\mu} + D_\mu^A\frac{\partial}{\partial y^A} + H_{\mu\nu}^A\frac{\partial}{\partial y_\nu^A}\right) \ ,    
\end{equation}
and the fact that this multivector field is holonomic ensures the {\sl{\sc sopde} condition}, $D^A_\mu = y^A_\mu$, is satisfied.

\item
Using a natural system of coordinates on $J^1\pi^*$, 
the section $j^1\phi$ satisfies the {\sl \textbf{Euler--Lagrange equations}} expressed as:
$$
\frac{\partial \mathscr{L}}{\partial y^A} \circ j^1\phi 
= \derpar{}{x^\mu} \left( \derpar{\mathscr{L}}{y^A_\mu} \circ j^1\phi \right)
\ .
$$
\een
\end{paragraph}

\begin{paragraph}{Singular Hamiltonian field theories:}
{\sl Singular Hamiltonian systems} in field theory are defined
on constraint submanifolds of $J^1\pi^*$.
In particular, let $\jmath_\circ\colon P_\circ\hookrightarrow J^1\pi^*$
be a $\tau$-transverse embedded submanifold such that
$P_\circ$ is a fibered submanifold over $E$ and $M$ with natural projections 
\beq
\tau_\circ:=\tau\circ\jmath_\circ\colon P_\circ\to E \quad , \quad
\overline{\tau}_\circ:=\pi\circ\tau_\circ\colon P_\circ\to M \ .
\label{projections}
\eeq
Now, if ${\rm h}\colon J^1\pi^*\to{\cal M}\pi$ is a section of $\sigma$, 
then ${\rm h}_\circ={\rm h}\circ\jmath_\circ\colon P_\circ\to{\cal M}\pi$
is called a {\sl Hamiltonian section} (for almost-regular theories)
and it allows us to define the following {\sl Hamilton--Cartan} forms on $P_\circ$:
$$
\Theta^\circ_{\mathscr H}={\rm h}_\circ^{\,*}\Theta\in\df^m(P_\circ)
\quad , \quad
\Omega^\circ_{\mathscr H}=-\d\Theta^\circ_{\mathscr H}={\rm h}_\circ^{\,*}\Omega\in\df^{m+1}(P_\circ) \ .
$$

The form $\Omega^\circ_{\mathscr H}$ is, in general, a premultisymplectic form
and thus the pair $(P_\circ,\Omega_{\mathscr{H}}^\circ)$
is a {\sl \textbf{ singular Hamiltonian system}}.
The Hamiltonian field equations are obtained from the variational principle for sections
$\Gamma(\overline{\tau}_\circ)$
similarly to the regular case.
The analog to the field equations \eqref{equationHEOM} read
\beq
\inn({\bf X}^\circ_\mathscr{H})\Omega^\circ_\mathscr{H}=0
\quad , \quad \mbox{\rm  for every ${\bf X}^\circ_{\mathscr{H}} \in  \{{\bf X}^\circ_{\mathscr{H}}\}
\subset\vf^m(P_\circ)$} \ ;
\label{dimgravmvfh}
\eeq
and the corresponding equation for the integral sections $\psi_\circ\colon M\to P_\circ$ of ${\bf X}^\circ_\mathscr{H}$ is
$$
\inn(\psi_\circ^{(m)})(\Omega_{\mathscr{H}}^\circ\circ\psi_\circ)=0\ .
$$
For soluble Hamiltonian systems, the Hamiltonian field equations have, in general, consistent solutions only on a $\overline{\tau}$-transverse embedded submanifold $P_f\hookrightarrow P_\circ$, called the {\sl final constraint submanifold},
obtained by using a suitable constraint algorithm
(see \cite{LMMMR-2005,GGR-2022}).
\end{paragraph}

For the convenience of the reader, we briefly summarize the constraint algorithm as follows:
\begin{itemize}
\item 
The field equations may yield some {\sl compatibility constraints} which specify the {\sl compatibility constraint submanifold} $P_1\hookrightarrow P_\circ$.

\item 
Impose {\sl stability (tangency)} of solutions to the variational problem by requiring that the Lie derivative of the compatibility constraint $P_1$ with respect to the vector fields components of $X^\circ_{\mathscr{H}}$ vanishes.
This may produce {\sl stability (tangency) constraints}. Perform the aforementioned Lie derivative operation on each new constraint which arises until the process terminates, which occurs when no new constraints are found. 
In the soluble cases, this geometric algorithm leaves us with a final constraint submanifold $P_f\hookrightarrow \dotsb \hookrightarrow P_1\hookrightarrow P_\circ$.
\end{itemize}

\paragraph{Singular Lagrangian field theories:}  
\label{LagHam}
We consider the particular case of Lagrangian field theories described by {\sl almost-regular Lagrangians}
for which the existence of an equivalent Hamiltonian formalism is assured (see \cite{LMM-96} for details).
A singular Lagrangian is {\sl \textbf{almost-regular}} if,
(i) $P_\circ=\text{Im}{\mathscr{F}\mathscr{L}}$ is a submanifold of $J^1\pi^*$,
(ii) ${\mathscr{FL}}$ is a submersion onto its image, and
(iii) for every ${\rm p}\in J^1\pi$, the fibers
${\mathscr{FL}}^{-1}({\mathscr{FL}}({\rm p}))$
are connected submanifolds of $J^1\pi$.

\begin{remark}
{\rm In \cite{LMM-96}, a stronger definition of almost-regularity is considered by demanding that:
(i) $\widetilde P_\circ=\text{Im}\widetilde{\mathscr F\mathscr{L}}$ is a submanifold of ${\cal M}\pi$,
(ii) $\widetilde{\mathscr F\mathscr{L}}$ is a submersion onto its image, and
(iii) for every ${\rm p}\in J^1\pi$, the fibers
${\mathscr{FL}}^{-1}({\mathscr{FL}}({\rm p}))$
and ${\widetilde{\mathscr{FL}}}^{-1}({\widetilde{\mathscr{FL}}}({\rm p}))$
are connected submanifolds of $J^1\pi$.
This definition implies ours (given in the previous paragraph); both definitions are locally equivalent.
Other alternative definitions are also considered in
\cite{GMS-97,Sa-95}.
}\end{remark}

For almost-regular Lagrangian systems,  
$\widetilde {P}_\circ=\text{Im}\widetilde{\mathscr{FL}}$
is a $\widetilde\sigma$-transverse submanifold 
with natural embedding $\widetilde\jmath_\circ\colon\widetilde P_\circ\hookrightarrow{\cal M}\pi$;
furthermore, $P_\circ=\text{Im}{\mathscr F\mathscr{L}}$
is a $\tau$-transverse submanifold 
with natural embedding $\jmath_\circ\colon P_\circ\hookrightarrow J^1\pi^*$,
and then we have the projections
\ref{projections}.
The restriction of the projection $\sigma$ to $\widetilde P_\circ$, denoted
$\widetilde\sigma_\circ\colon\widetilde P_\circ\to P_\circ$, is a (local) diffeomorphism.
Taking $\widetilde{\rm h}:=\widetilde\sigma_\circ^{-1}$
gives a Hamiltonian section ${\rm h}_\circ:=\widetilde\jmath_\circ\circ\widetilde{\rm h}$.
The corresponding Hamilton--Cartan forms
$\Theta^\circ_{\mathscr H}={\rm h}_\circ^*\Theta$ and $\Omega^\circ_{\mathscr H}=-\d\Theta^\circ_{\mathscr H}={\rm h}_\circ^*\Omega$
verify that 
${\mathscr F}\mathscr{L}_\circ^*\Theta^\circ_{\mathscr H}=\Theta_{\Lag}$ and
${\mathscr F}\mathscr{L}_\circ^*\Omega^\circ_{\mathscr H}=\Omega_{\Lag}$,
where ${\mathscr F}\mathscr{L}_\circ$ is the restriction map of
${\mathscr F}\mathscr{L}$ onto $P_\circ$, given as
${\mathscr F}\mathscr{L}=\jmath_\circ\circ{\mathscr F}\mathscr{L}_\circ$.
Furthermore, there exists a De Donder--Weyl Hamiltonian
function $\mathscr{H}_\circ\in C^\infty(P_\circ)$
such that
$E_\mathscr{L}=\mathscr{FL}_\circ^*\mathscr{H}_\circ$ and the pair $(P_\circ,\Omega_\mathscr{H}^\circ)$
is the {\sl Hamiltonian system associated with the singular Lagrangian system} $(J^1\pi,\Omega_\L)$.
In order to define the De Donder--Weyl Hamiltonian
function and the energy function globally, it is necessary to use a connection on the bundle $\pi\colon E\to M$ \cite{CCI-91,EMR-96,EMR-99b}. 
However, the De Donder--Weyl Hamiltonian
function can, in general, be defined locally without using a fiber bundle connection; locally, the trivial connection induced on the local chart is used.
The Lagrangian field equations are 
satisfied by critical sections, $\Gamma(\overline{\pi}^1)$,
which solve the variational problem as in the regular case.
In soluble Lagrangian systems, the Lagrangian field equations have, in general, consistent solutions only on a $\overline{\pi}^1$-transverse embedded submanifold $S_f\hookrightarrow J^1\pi$, the {\sl final Lagrangian constraint submanifold}
obtained by using the constraint algorithm (see \cite{AGR-2022,LMMMR-2005,GGR-2022}).

As above, we briefly summarize the constraint algorithm as follows:
\begin{itemize}
\item 
The field equations may yield some {\sl compatibility constraints} which specify the {\sl compatibility constraint submanifold} $S_1\hookrightarrow J^1\pi$.
\item 
Impose the {\sc sopde} condition $D^A_\mu = y^A_\mu$ in the local expression \eqref{XL} of ${\bf X}_\mathscr{L}$.
This may produce a {\sl {\sc sopde} constraint submanifold} $S_2\hookrightarrow S_1\hookrightarrow J^1\pi$.
\item 
Impose {\sl stability (tangency)} of solutions to the variational problem by requiring that the Lie derivative of 
any constraint found in the previous steps
with respect to the vector fields components of ${\bf X}_\mathscr{L}$ vanishes.
This may produce {\sl stability (tangency) constraint}. Perform the aforementioned Lie derivative operation on each new constraint which arises until no new constraints are found
and the process terminates. 
In soluble cases, this geometric algorithm leaves us with a final constraint submanifold $S_f\hookrightarrow \dotsb S_2 \hookrightarrow S_1\hookrightarrow J^1\pi$.
\end{itemize}

The geometric structure of singular field theories is illustrated in the following diagram:
 \begin{center}
    \begin{tikzcd}[column sep=1.75cm, row sep=1.25cm]
       & \widetilde P_\circ \arrow[d, "\widetilde\sigma_\circ"] \arrow[rr, "\widetilde\jmath_\circ"] & & \mathcal{M}\pi \arrow[d, swap, "\sigma"] \\
       J^1\pi \arrow[ur, "\widetilde{\mathscr{FL}_\circ}"] \arrow[r, swap, "\mathscr{FL}_\circ"] \arrow[ddrr, swap, "\overline{\pi}^1"] & P_\circ \arrow[u, bend left=30, pos=0.49, "\widetilde{\mathrm{h}}"] \arrow[rru, "\mathrm{h}_\circ"] \arrow[ddr, swap, "\overline{\tau}_\circ"] \arrow[dr, "\tau_\circ"] \arrow[rr, swap, "\jmath_\circ"] & & J^1\pi^\ast \arrow[u, swap, bend right=30, pos=0.49, "\mathrm{h}"] \arrow[dl, swap, "\tau"] \arrow[ddl, "\overline{\tau}"] \\
       & & E \arrow[d, "\pi"] \\
       & & M
    \end{tikzcd}
\end{center}
(See also diagrams \eqref{diagram1}
and \eqref{diagram2}.)

\subsection{Lifting transformations and vector fields onto the configuration bundle}
\label{lifting}

Classical field theories 
often exhibit symmetries that are associated with diffeomorphisms on the base manifold $M$
(e.g., spacetime diffeomorphisms).
In order to lift these base space transformations (on $M$) to the Hamiltonian multiphase space $J^1\pi^*$, we must first discuss how to lift such diffeomorphisms on $M$ to the configuration bundle $\pi\colon E\to M$ (see \cite{GGR-2022,GR-2023}).
Note that it is not always the case that canonical lifts exist from $M$ to $E$; for such lifts to exist, it is necessary that $E$ is a bundle with some additional natural structure. 
For instance, as in Yang--Mills and in several other relevant bosonic field theories in physics, $E$ is a tensor bundle over $M$ for which it is well-known that the canonical lifts from $M$ to $E$ exist;
as another example, such as in Einstein--Cartan theory discussed in Section \ref{Einstein-Cartan}, $E$ is the bundle of frames and spin connections which is also natural.

Consider a vector field $\displaystyle \xi_M=-\xi^\mu(x)\frac{\partial}{\partial x^\mu}\in\vf(M)$
which generates infinitesimal diffeomorphisms $\Phi_M\colon M\rightarrow M$, 
i.e., infinitesimal coordinate transformations $x'^\mu=x^\mu+\xi^\mu(x)$;
such diffeomorphisms, the one-parameter subgroup of local diffeomorphisms which are generated infinitesimally by some vector field $\xi_M$ are the only kind that we will be interested in.
A generic vector field on $E$ which projects to $\xi_M$ has the coordinate expression
\begin{equation}
\label{diffsEvector}
\xi_E=-\xi^\mu(x)\frac{\partial}{\partial x^\mu}-\xi^A(x,y)\frac{\partial}{\partial y^A}\in\vf(E)\ ,
\end{equation}
and generates the infinitesimal transformations
$(x^\mu,y^A)\rightarrow (x^\mu+\xi^\mu(x),y^A+\xi^A(x,y))$ on $E$,
with $\xi^A(x,y)\in C^\infty(E)$.
Recall that the fields $y^A(x)$ are components of the sections $\phi(x)=(x^\mu,y^A(x))$ of $\pi\colon E\to M$ 
and transform as the Lie derivatives with respect to $\xi_M$; that is, 
\begin{equation}
\label{liederivative}
\delta y^A(x)\equiv\Lie(\xi_M) y^A(x)=y'^A(x)-y^A(x)=-\xi^\mu(x)\derpar{y^A}{x^\mu}(x) +\widetilde{\xi}^A(x)\ .
\end{equation}
These field transformations are referred to as the {\sl \textbf{local variation}} of the fields.
The term  $\displaystyle -\xi^\mu(x)\partial_\mu y^A$ 
in (\ref{liederivative}) is known as the {\sl \textbf{transport term}} 
and $\widetilde{\xi}^A(x)$ are called the {\sl \textbf{global variation}} of the fields, given by
\begin{equation*}
\widetilde{\xi}^A(x)= y'^A(x')-y^A(x)\ .
\end{equation*}
Note that in this notation, $y'^A(x)$ is a different point on along the fiber over the spacetime point $x$. 
Furthermore, $y'^A(x')\rightarrow y'^A(x)$ as $x'^\mu \rightarrow x^\mu$.

One can interpret the local field variations as follows 
\cite{GIMMSY,GP-2002,KSM-2011,Mi-2008}:

\begin{definition}
Consider a section $\phi\colon M\to E:x^\mu\to (x^\mu,\phi^A(x))$, and let $\xi_E\in\vf(E)$ be a $\pi$-projectable vector field whose local expression is \eqref{diffsEvector}, which projects to a vector field $\displaystyle \xi_M=-\xi^\mu(x)\frac{\partial}{\partial x^\mu}\in\vf(M)$
which generates infinitesimal diffeomorphisms $\Phi_M\colon M\rightarrow M$.
The \textbf{generalized Lie derivative of the section $\phi$ by $\xi_M$} is the map
 $\mathbb{L}(\xi_M)\phi\colon M\to\Tan E$
  defined as
\begin{equation}
\label{LieDerivSections1}
\mathbb{L}(\xi_M)\phi=\Tan\phi\circ \xi_M-\xi_E\circ\phi \ ,
\end{equation}
which is a vector field along $\phi$.
This generalized Lie derivative has the form 
$\mathbb{L}(\xi_M)\phi=(\phi,\Lie(\xi_M)\phi)$,
where the section $\Lie(\xi_M)\phi\colon M\to V(\Tan \pi)$ is called the
\textbf{Lie derivative of the section $\phi$ by $\xi_M$}, and is expressed as 
$\displaystyle \mathbb{L}(\xi_M)\phi=\Big( x^\mu,\xi^\nu\,\partial_\nu\phi^A-\xi^A\circ\phi\Big)$, 
given the local expression \eqref{diffsEvector} of $\xi_E$.
\end{definition}

The above definition of the Lie derivative \eqref{LieDerivSections1} is in agreement with the Lie derivative of $y^A(x)$ 
defined by the field transformations $\delta y^A(x)$ in \eqref{liederivative} if we set
$\Lie(\xi_M)\phi^A=\Lie(\xi_M)y^A(x)$.
Then, the functions $\widetilde{\xi}^A(x)$ in \eqref{liederivative} are given in terms of the component functions $\xi^A(x,y)$ of $\xi_E$ and any local section $\phi$ as 
\begin{equation}
\label{xiA}
\widetilde{\xi}^A(x)=\xi^A(x,y)\circ\phi\ .
\end{equation}
For theories in which $y^A(x)$ are scalar fields, diffeomorphisms on the base space $M$ produce field variations $\delta y^A(x)$ for which $\widetilde{\xi}^A(x)=0$, and then $\xi^A(x,y)=0$.

A particular situation occurs when the fields $y^A(x)$ are tensor fields,
$T\equiv (T^{\mu_1,\dotsc,\mu_r}_{\nu_1,\dotsc,\nu_s}(x))\in\mathfrak{T}^{(k,l)}(\Tan M)$,
of type $(k,l)\not=(0,0)$ on $M$
so that $E$ is a tensor bundle over $M$. 
Then, the functions $\widetilde{\xi}^A(x)$ are obtained from the induced tensor transformation laws 
in which the Jacobian matrices associated with the coordinate transformations $x^\mu\rightarrow x^\mu+\xi^\mu(x)$
are used to transform the indices of the tensor field
as usual
(see e.g. \cite{Li-2016}):
\begin{equation}
\label{TensorTransformation}
(T')^{\mu_1,\dotsc,\mu_r}_{\nu_1,\dotsc,\nu_s}(x') = \left(\frac{\partial x'^{\mu_1}}{\partial x^{\alpha_1}}\right) \dotsb \left(\frac{\partial x'^{\mu_r}}{\partial x^{\alpha_r}}\right) \left(\frac{\partial x^{\beta_1}}{\partial x'^{\nu_1}}\right) \dotsb \left(\frac{\partial x^{\beta_s}}{\partial x'^{\nu_s}}\right) T^{\alpha_1,\dotsc,\alpha_r}_{\beta_1,\dotsc,\beta_s}(x)\ .
\end{equation}
The Taylor expansion, to first order, of the left-hand side of the equation above around $x$ gives the transport term as
\beq
(T')^{\mu_1,\dotsc,\mu_r}_{\nu_1,\dotsc,\nu_s} (x') = (T')^{\mu_1,\dotsc,\mu_r}_{\nu_1,\dotsc,\nu_s} (x+\xi)
= (T')^{\mu_1,\dotsc,\mu_r}_{\nu_1,\dotsc,\nu_s} (x) + \xi^\lambda\,\derpar{(T')^{\mu_1,\dotsc,\mu_r}_{\nu_1,\dotsc,\nu_s}(x)}{x^\lambda} \ ,
\label{tensorchange}
\eeq
while the Taylor expansion of the Jacobian matrices on the right-hand side of equation \eqref{tensorchange}
gives the global variation of the fields $\Delta T^{\mu_1,\dotsc,\mu_r}_{\nu_1,\dotsc,\nu_s}(x)$ as
\begin{align}
\label{DeltaT}
\Delta T^{\mu_1,\dotsc,\mu_r}_{\nu_1,\dotsc,\nu_s}= \sum_{\mu=\mu_1}^{\mu_r}\Big(\derpar{\xi^\mu}{x^\lambda}\Big)T^{\mu_1,\dotsc,(\mu\rightarrow\lambda),\dotsc,\mu_r}_{\nu_1,\dotsc,\nu_s} -\sum_{\nu=\nu_1}^{\nu_s}\Big(\derpar{\xi^\lambda}{x^\nu}\Big)T^{\mu_1,\dotsc,\mu_r}_{\nu_1,\dotsc,(\nu\rightarrow \lambda),\dotsc,\nu_s}\ .
\end{align} 
Then, the local field variation given by the Lie derivative with respect to $\displaystyle \xi_M\in\vf(M)$ is
$$
\delta T^{\mu_1,\dotsc,\mu_r}_{\nu_1,\dotsc,\nu_s}(x)=
(T')^{\mu_1,\dotsc,\mu_r}_{\nu_1,\dotsc,\nu_s}(x) - T^{\mu_1,\dotsc,\mu_r}_{\nu_1,\dotsc,\nu_s}(x)=  
\Lie(Z)T^{\mu_1,\dotsc,\mu_r}_{\nu_1,\dotsc,\nu_s}=
-\xi^\lambda\,\derpar{(T')^{\mu_1,\dotsc,\mu_r}_{\nu_1,\dotsc,\nu_s}}{x^\lambda}+\Delta T^{\mu_1,\dotsc,\mu_r}_{\nu_1,\dotsc,\nu_s}\ ;
$$
it follows from \eqref{xiA} that,
\begin{equation}
\label{Tensor_xiE}
\xi_E=- \xi^\mu\derpar{}{x^\mu}-\Delta^{\mu_1,\dotsc,\mu_r}_{\nu_1,\dotsc,\nu_s}(x,T)\,\frac{\partial}{\partial T^{\mu_1,\dotsc,\mu_r}_{\nu_1,\dotsc,\nu_s}}\ ,
\end{equation}
where,
\begin{equation}
\label{DeltaT2}
\Delta T^{\mu_1,\dotsc,\mu_r}_{\nu_1,\dotsc,\nu_s}(x)=\Delta^{\mu_1,\dotsc,\mu_r}_{\nu_1,\dotsc,\nu_s}(x,T)\circ\phi\ .
\end{equation}
It is important to point out that, when $E$ is a tensor bundle over $M$ with coordinates $(x^\mu,T^{\mu_1,\dotsc,\mu_r}_{\nu_1,\dotsc,\nu_s})$,
then \eqref{Tensor_xiE} is just the canonical lift of $\displaystyle \xi_M\in\vf(M)$ to $E$ 
which is defined as:

\begin{definition}
\ben
\item
Let $\Phi_M\colon M\to M$ be a diffeomorphism.
The \textbf{canonical lift of $\Phi_M$ to $E$} 
is the diffeomorphism $\Phi_E\colon E\to E$ 
such that, for every $(x,T_x)\in E$,
with $T_x\in\mathfrak{T}^{(k,l)}(\Tan_xM)$,
$$
\Phi_E(x,T_x):=(\Phi_M(x),{\cal T}\Phi_M(T_x))\ ,
$$
where ${\cal T}\Phi_M$ denotes the canonical transformation of tensors on $M$ induced by $\Phi_M$.
Thus, $\pi\circ\Phi_E=\Phi_M\circ\pi$.
\item
Let $\xi_M\in\vf (M)$ be a vector field whose flux is made of the local one-parameter groups of diffeomorphisms of $M$, denoted $\phi_t$.
The \textbf{canonical lift of $\xi_M$ to $E$}
is the vector field $\xi_E\in\vf(E)$ whose flux is made of the
local one-parameter groups of diffeomorphisms  $(\phi_{_E})_t$
which are the canonical lifts of $\phi_t$ to $E$.
\een
\label{lift}
\end{definition}

Thus, the definition of the Lie derivative \eqref{LieDerivSections1} of local sections $\phi$ is stated to agree with the Lie derivative of fields \eqref{liederivative}, 
where $\xi_E$ is the canonical lift of $\xi_M$ to $E$ and, hence, the relation \eqref{xiA} holds.

\section{Canonical lifts on the multimomentum bundles}

In this section we present multimomentum bundle diffeomorphisms $\Phi_{\mathcal{M}\pi}:\mathcal{M}\pi\to \mathcal{M}\pi$ and $\Phi_{J^1\pi^*}:J^1\pi^*\to J^1\pi^*$ as {\sl canonical lifts} which are fiber preserving over the configuration manifold $E$, as depicted in the following diagram: 
 \begin{equation}\label{LiftDiag}
    \begin{tikzcd}[column sep=2cm, row sep=1.75cm]
       \mathcal{M}\pi \arrow[r, "\Phi_{\mathcal{M}\pi}"] \arrow[d, "\sigma"] \arrow[dd, bend right=30, swap, "\widetilde\sigma"] & \mathcal{M}\pi  \arrow[d, swap, "\sigma"] \arrow[dd, bend left=30, "\widetilde\sigma"]  \\
       J^1\pi^\ast \arrow[r, "\Phi_{J^1\pi^\ast}"] \arrow[d, "\tau"] & J^1\pi^* \arrow[d, swap, "\tau"]\\
       E \arrow[r,"\Phi_E"] & E
    \end{tikzcd}
\end{equation}
This allows us to define the notion of {\sl natural symmetries}
of Hamiltonian systems of the De Donder--Weyl type in classical field theory.
We also present {\sl canonical lifts} from $E$ to embedded constraint submanifolds of ${\cal M}\pi$ and $J^1\pi^*$, relevant to the study of singular field theories.

\subsection{Lifting diffeomorphisms and vector fields to the bundle ${\cal M}\pi$}
\label{lifting1}

Consider a regular Hamiltonian system $(J^1\pi^*,\Omega_{\mathscr{H}})$.
As ${\cal M}\pi\equiv\Lambda_2^m\Tan^*E$ is a bundle of forms over $E$,
the canonical lift of diffeomorphisms from $E$ to ${\cal M}\pi$
is carried out in a natural way 
as is described for tensor bundles in the above Definition \ref{lift}:

\begin{definition}
\label{liftdifeotilde}
Let $\Phi_E\colon E\to E$ be a diffeomorphism on $E$ which induces a
diffeomorphism $\Phi_M\colon M\to M$, on $M$;
that is, $\Phi_M\circ\pi=\pi\circ\Phi_E$.
The \textbf{canonical lift of $\Phi_E$ to ${\cal M}\pi$}
is the diffeomorphism $\Phi_{{\cal M}\pi}\colon{\cal M}\pi\to{\cal M}\pi$ 
such that, for every ${\rm p}=(y,\zeta)\in{\cal M}\pi$,
with $\zeta\in\Lambda_2^m\Tan_y^*E$,
$$
\Phi_{{\cal M}\pi}(y,\zeta):=(\Phi_E(y),(\Phi_E^{\,-1})_y^*\,\zeta) \ .
$$
\end{definition}

Such $\Phi_{{\cal M}\pi}$ are $\widetilde\sigma$-fiber preserving diffeomorphisms:
$\Phi_E\circ\widetilde\sigma=\widetilde\sigma\circ\Phi_{{\cal M}\pi}$ (see Diagram \eqref{LiftDiag}).
Note that we will only be interested in lifting the one-parameter subgroup of local diffeomorphisms of $E$ which are generated infinitesimally by some vector field $\xi_E\in\vf(E)$, those which in turn are the lifts of from $M$ to $E$ as discussed in Section \ref{lifting}.

These canonical lifts $\Phi_{{\cal M}\pi}$
preserve the canonical structures of ${\cal M}\pi$ which consist of
the tautological form $\Theta\in\df^m({\cal M}\pi)$ and
the multisymplectic form $\Omega\in\df^{m+1}({\cal M}\pi)$. In fact:

\begin{proposition} 
If $\Phi_E\colon E\to E$ is a diffeomorphism
which induces a
diffeomorphism $\Phi_M\colon M\to M$,
and $\Phi_{{\cal M}\pi}$ is its canonical lift to ${\cal M}\pi$, then 
$\Phi_{{\cal M}\pi}^{\ *}\,\Theta =\Theta$ and
$\Phi_{{\cal M}\pi}^{\ *}\,\Omega =\Omega$.
\label{levdif}
\end{proposition}
\proof
For every $y\in E$, $\zeta\in\Lambda_2^m\Tan_y^*E$,
and $X_0,\ldots,X_{m-1}\in\Tan_{(y,\zeta)}({\cal M}\pi)$,
using the definition of $\Theta$ given in \eqref{Thetadef}, and the property
$\Phi_E^{-1}\circ\widetilde\sigma\circ\Phi_{{\cal M}\pi}=\widetilde\sigma$, we obtain
\beann
(\Phi_{{\cal M}\pi}^{\ *}\Theta)_{(y,\zeta)}(X_0,\ldots,X_{m-1})=
\Theta_{(\Phi_E(y),(\Phi_E^{-1})^*\zeta)}\Big(\Tan_{(y,\zeta)}\Phi_{{\cal M}\pi}(X_0),\ldots,\Tan_{(y,\zeta)}\Phi_{{\cal M}\pi}(X_{m-1})\Big)=
\\ 
\big((\Phi_E^{-1})_{\Phi_E(y)}^{\,*}\zeta\big)\Big(\Tan_{(\Phi_E(y),(\Phi_E^{-1})^*\zeta)}\widetilde\sigma\big(\Tan_{(y,\zeta)}\Phi_{{\cal M}\pi}(X_0)\big),\ldots,\Tan_{(\Phi_E(y),(\Phi_E^{-1})^*\zeta)}\widetilde\sigma\big(\Tan_{(y,\zeta)}\Phi_{{\cal M}\pi}(X_{m-1})\big)\Big)=
\\ 
\big((\Phi_E^{-1})_{\Phi_E(y)}^{\,*}\zeta\big)\Big(\Tan_{(y,\zeta)}(\widetilde\sigma\circ\Phi_{{\cal M}\pi})(X_0),\ldots,\Tan_{(y,\zeta)}(\widetilde\sigma\circ\Phi_{{\cal M}\pi})(X_{m-1})\Big)=
\\ 
\zeta\Big(\Tan_{\Phi_E(y)}\Phi_E^{-1}\big(\Tan_{(y,\zeta)}(\widetilde\sigma\circ\Phi_{{\cal M}\pi})(X_0)\big),\ldots,\Tan_{y}\Phi_E^{-1}\big(\Tan_{(y,\zeta)}(\widetilde\sigma\circ\Phi_{{\cal M}\pi})(X_{m-1})\big)\Big)=
\\ 
\zeta\Big(\Tan_{(y,\zeta)}(\Phi_E^{-1}\circ\widetilde\sigma\circ\Phi_{{\cal M}\pi})(X_0),\ldots,\Tan_{(y,\zeta)}(\Phi_E^{-1}\circ\widetilde\sigma\circ\Phi_{{\cal M}\pi})(X_{m-1})\Big)=
\\ 
\zeta\Big(\Tan_{(y,\zeta)}\widetilde\sigma(X_0),\dotsc ,\Tan_{(y,\zeta)}\widetilde\sigma(X_{m-1})\Big)=
\Theta_{(y,\zeta)}(X_0,\ldots,X_{m-1}) \ , \quad
\eeann
hence $\Phi_{{\cal M}\pi}^{\ *}\,\Theta =\Theta$.
The result for $\Omega$ is immediate.
\qed

Now, if $\xi_E\in\vf(E)$ is a $\pi$-projectable vector field (i.e.; there exist $\xi_M\in\vf(M)$ 
such that the local flows of $\xi_M$ and $\xi_E$ are $\pi$-related),
then it is possible to define the canonical lift of $\xi_E$ to ${\cal M}\pi$ as:

\begin{definition}
Let $\xi_E\in\vf(E)$ be a $\pi$-projectable vector field whose flux is made of local one-parameter groups of diffeomorphisms of $E$, denoted $\phi_t$.
The \textbf{canonical lift of $\xi_E$ to ${\cal M}\pi$}
is the vector field $Z_\xi\in\vf({\cal M}\pi)$ whose flux is made of the
local one-parameter groups of diffeomorphisms  $\widetilde\phi_t$
which are the canonical lifts of $\phi_t$ to ${\cal M}\pi$.
\end{definition}


As a straightforward consequence of the previous definition
and Proposition \ref{levdif}, it follows that

\begin{proposition}
\label{liftproject}
The canonical lift, $Z_\xi\in{\cal M}\pi$, of a $\pi$-projectable vector field $\xi_E\in\vf(E)$ has the following properties:
\ben
\item
$Z_\xi$ is $\widetilde\sigma$-projectable and $\widetilde\sigma_*Z_\xi=\xi_E$\ .
\item 
$\Lie(Z_\xi)\Theta=0$ and $\Lie(Z_\xi)\Omega=0$\ .
\een
\end{proposition}

Such vector field lifts, $Z_\xi\in\vf({\cal M}\pi)$,
can be characterized using the canonical forms of ${\cal M}\pi$. First we define:

\begin{definition}
Every vector field $\xi_E\in\vf(E)$
induces a form $\Gamma_\xi\in\df^{m-1}({\cal M}\pi)$ defined as follows:
for every point $(y,\zeta)\in{\cal M}\pi$, 
$$
(\Gamma_\xi)_{(y,\zeta)}:=i(\xi_E\vert_y)\zeta \ .
$$
\end{definition}

As an immediate consequence of this definition:

\begin{proposition}
\label{formzeta}
If $\xi_E\in\vf(E)$ is a $\pi$-projectable vector field and
$Z_\xi\in\vf({\cal M}\pi)$ is
the canonical lift of $\xi_E$ to ${\cal M}\pi$, then
$$
\Gamma_\xi=i(Z_\xi)\Theta \ .
$$
\end{proposition}
\proof
The $\widetilde\sigma$-projectivity property of the vector field $Z_\xi$ means that,
for every $(y,\zeta)\in{\cal M}\pi$,
it follows that $\Tan_{(y,\zeta)}\widetilde\sigma(Z_\xi\vert_{(y,\zeta)})=\xi_E\vert_y$.
Then, taking into account the definition of the canonical $m$-form $\Theta$,
it follows that
$$
(\Gamma_\xi)_{(y,\zeta)}=\zeta(\xi_E\vert_y)=\zeta \big(\Tan_{(y,\zeta)}\widetilde\sigma(Z_\xi\vert_{(y,\zeta)})\big)=\Theta_{(y,\zeta)}(Z_\xi\vert_{(y,\zeta)}) \ ,
$$
and the statement holds.
\qed

From this proposition, we obtain:

\begin{proposition}
If $\xi_E\in\vf(E)$ is a $\pi$-projectable vector field, then the canonical lift of $\xi_E$ to ${\cal M}\pi$
is the unique vector field $Z_\xi\in\vf({\cal M}\pi)$ such that,
$$
\inn (Z_\xi)\Omega =\d\Gamma_\xi \ .
$$
That is,
$Z_\xi$ is the  (global) Hamiltonian vector field associated with the $(m-1)$-form $\Gamma_\xi$.
\label{hamlev}
\end{proposition}
\proof
Taking into account that 
$\Lie(Z_\xi)\Theta=0$, we obtain
$$
\inn(Z_\xi)\Omega =-\inn (Z_\xi)\d \Theta =
\d\inn(Z_\xi)\Theta -\Lie (Z_\xi)\Theta=\d(\Theta (Z_\xi))=\d\Gamma_\xi \ .
$$
Furthermore, given the local coordinate expression of a $\pi$-projectable vector field
$\xi_E\in\vf(E)$,
\beq
\label{Ycoor}
\xi_E=-\xi^\mu(x^\nu)\derpar{}{x^\mu}-\xi^A(x^\nu,y^B)\frac{\partial}{\partial y^A}\ ,
\eeq
then,
\beann
\Gamma_\xi&=&\xi^\nu p_A^\mu\,\d y^A\wedge\d ^{m-2}x_{\mu\nu}-
\left(\xi^A p_A^\mu+\xi^\mu p\right)\,\d^{m-1}x_\mu \ , 
\\
\d\Gamma_\xi&=&\xi^\nu\,\d p_A^\mu\,\d y^A\wedge\d ^{m-2}x_{\mu\nu}+
\left(\derpar{\xi^\mu}{x^\nu} p_A^\nu-\derpar{\xi^\nu}{x^\nu} p_A^\mu-\derpar{\xi^B}{y^A}p_B^\mu\right)\d y^A\wedge\d ^{m-1}x_{\mu}
\\ & &
-\xi^A\,\d p_A^\mu\wedge\d^{m-1}x_\mu-\xi^\mu \,\d p\wedge\d^{m-1}x_\mu
-\left(\derpar{\xi^A}{x^\mu}p_A^\mu+\derpar{\xi^\mu}{x^\mu} p\right)\d^mx \ .
\eeann
Using Proposition \ref{hamlev}
and equation \eqref{coordforms},
it follows that the canonical lift of $\xi_E$ to ${\cal M}\pi$ is uniquely given as
\begin{equation}
\label{liftJ1}
Z_\xi=-\xi^\mu\derpar{}{x^\mu}-\xi^A\frac{\partial}{\partial y^A}-\left(\derpar{\xi^\mu}{x^\nu} p_A^\nu-\derpar{\xi^\nu}{x^\nu} p_A^\mu-\derpar{\xi^B}{y^A}p_B^\mu\right)\frac{\partial}{\partial p_A^\mu}+\left(\derpar{\xi^A}{x^\mu}p^\mu_A+\derpar{\xi^\mu}{x^\mu}p\right)\derpar{}{p} \ . 
\end{equation}
\qed

\begin{remark}
{\rm
It is important to point out that the existence of a unique Hamiltonian vector field
associated with the $(m-1)$-form $\Gamma_\xi$
is assured as a consequence of the assumption
that $\xi_E$ is a $\pi$-projectable vector field;
this is also shown in the above calculations in local coordinates.
}    
\end{remark}

This expression shows clearly that, as it is stated in item 1 of Proposition \ref{liftproject},
$Z_\xi$ is $\widetilde\sigma$-projectable and $\widetilde\sigma_*Z_\xi=\xi_E$.

\subsection{Lifting diffeomorphisms and vector fields to the multimomentum bundle $J^1\pi^*$}
\label{lifting3}

If $\Phi_E\colon E\to E$ is a diffeomorphism on $E$ which induces a
diffeomorphism $\Phi_M\colon M\to M$, then the induced map $\Phi_E^*\colon\df^m(E)\to\df^m(E)$
transforms $\pi$-semibasic forms into themselves.
Hence, bearing in mind the definition of
$J^1\pi^*:={\cal M}\pi/\Lambda^m_1(\Tan^*{\rm E})$ and Definition \ref{liftdifeotilde},
this implies that the canonical lifts
$\Phi_{{\cal M}\pi}:{\cal M}\pi\to{\cal M}\pi$ induce diffeomorphisms on $J^1\pi^*$,
leading to the following definition:

\begin{definition}
\label{liftdifeohat}
Let $\Phi_E\colon E\to E$ be a diffeomorphism on $E$ which induces a diffeomorphism $\Phi_M\colon M\to M$ on $M$ and $\Phi_{{\cal M}\pi}\colon{\cal M}\pi\to{\cal M}\pi$ is the canonical lift of $\Phi_E$ to ${\cal M}\pi$.
The \textbf{canonical lift of $\Phi_E$ to $J^1\pi^*$} is the diffeomorphism $\Phi_{J^1\pi^*}\colon J^1\pi^*\to J^1\pi^*$
induced on $J^1\pi^*$ by $\Phi_{{\cal M}\pi}$; that is, 
$$ 
\Phi_{J^1\pi^*}\circ\sigma=\sigma\circ\Phi_{{\cal M}\pi} \ . 
$$  
\end{definition}

As $\Phi_E\circ\widetilde\sigma=\widetilde\sigma\circ\Phi_{{\cal M}\pi}$,
this also implies that
$\Phi_E\circ\widetilde\sigma=\widetilde\sigma\circ\Phi_{J^1\pi^*}$ (recall Diagram \eqref{LiftDiag}).

\begin{remark}
{\rm It is interesting to point out that $J^1\pi^*$ is canonically diffeomorphic to
$\pi^*\Tan M\otimes_E V^*E\otimes_E\pi^*\Lambda^m\Tan^*M$ (see, for instance, \cite{EMR-00b}).
Therefore, for a diffeomorphism $\Phi_E\colon E\to E$ 
which restricts to a diffeomorphism $\Phi_M\colon M\to M$, 
the definition of the canonical lift of $\Phi_E$ to $J^1\pi^*$
can be stated straightforwardly in a natural way,
using the same pattern as in Definition \ref{liftdifeotilde}.
}\end{remark}

Then, as on ${\cal M}\pi$, we can define:

\begin{definition} 
\label{liftvfhat}
Let $\xi_E\in\vf(E)$ be a $\pi$-projectable vector field whose flux is made of local one-parameter groups of diffeomorphisms of $E$, denoted $\phi_t$. 
The \textbf{canonical lift of $\xi_E$ to $J^1\pi^*$} is the vector field $Y_\xi\in\vf(J^1\pi^*)$ whose flux is made of the local one-parameter groups of diffeomorphisms $\widehat\phi_t$ which are the canonical lifts of $\phi_t$ to $J^1\pi^*$. 
\end{definition}

It follows from the local expression \eqref{liftJ1} that
canonical lifts $Z_\xi\in{\cal M}\pi$ are $\sigma$-projectable vector fields.
Therefore, from the above definitions, we conclude the following:

\begin{proposition}
$\sigma_*Z_\xi=Y_\xi$.
\end{proposition}

Additionally, by Proposition \ref{liftproject} it follows that
$\widetilde\sigma_*Z_\xi=\xi_E$, and therefore,
$$
\xi_E=\widetilde\sigma_*Z_\xi=(\tau\circ\sigma)_*Z_\xi=\tau_*\big(\sigma_*Z_\xi\big)=\tau_*Y_\xi \ .
$$

Then, the coordinate expression of the canonical lift $Y_\xi$ is,
\begin{equation}
\label{liftJ1b}
Y_\xi=-\xi^\mu\derpar{}{x^\mu}-\xi^A\frac{\partial}{\partial y^A}-\left(\derpar{\xi^\mu}{x^\nu} p_A^\nu-\derpar{\xi^\nu}{x^\nu} p_A^\mu-\derpar{\xi^B}{y^A}p_B^\mu\right)\frac{\partial}{\partial p_A^\mu}\ .
\end{equation}

Now, let $(J^1\pi^*,\Omega_{\mathscr{H}})$ be
the Hamiltonian system associated with a (hyper-)regular Lagrangian system
$(J^1\pi,\Omega_\mathscr{L})$ (as described in paragraph \ref{canreg}).
If $X_\xi=j^1\xi_E\in\vf(J
\pi)$ is the canonical lift to $J^1\pi$ of a $\pi$-projectable vector field $\xi_E\in\vf(E)$ whose local expression is given in equation \eqref{Ycoor}, then, the local expression for $X_\xi$ is given as (see \cite{EMR-96,book:Saunders89}),
\begin{equation}
\label{j1Y}
X_\xi=j^1\xi_E=-\xi^\mu\derpar{}{x^\mu}-\xi^A\frac{\partial}{\partial y^A}-
\left(\derpar{\xi^A}{x^\mu}-y^A_\nu\derpar{\xi^\nu}{x^\mu}+y^B_\mu\frac{\partial \xi^A}{\partial y^B}\right)\frac{\partial}{\partial y^A_\mu}\ .
\end{equation}
It thereby follows that:

\begin{proposition}
\label{propXYequivalence}
If the Lagrangian $\mathscr{L}$ is (hyper-)regular and $\Lie(X_\xi){\bf L}=0$, then $\mathscr{FL}_*X_\xi=Y_\xi$.
\label{FLproj}
\end{proposition}
\proof
As the Lagrangian $\mathscr{L}$ is (hyper-)regular, $\mathscr{FL}$ is a (global) diffeomorphism,
and the matrix of the tangent map $\mathscr{FL}_*$ in a natural coordinate system on $J^1\pi$ is given as
$$
 \left(\begin{matrix}{\rm Id} & 0 & 0 \cr 
 0 & {\rm Id} & 0 \cr
 \displaystyle\frac{\partial^2\mathscr{L}}{\partial x^\mu\partial y^B_\nu} &
 \displaystyle\frac{\partial^2\mathscr{L}}{\partial y^A\partial y^B_\nu} &
 \displaystyle\frac{\partial^2\mathscr{L}}{\partial y^A_\mu\partial y^B_\nu}\cr\end{matrix}\right) \ .
$$
Then, if 
\begin{equation*}
\mathscr{FL}_*X_\xi=f^\mu\derpar{}{x^\mu}+f^A\frac{\partial}{\partial y^A}+ f^\nu_B \frac{\partial}{\partial p^B_\nu} \ ,
\end{equation*}
while bearing in mind \eqref{j1Y}, then $f^\mu=-\xi^\mu(x^\nu)$, $f^A=-\xi^A(x^\nu,y^B)$, and
\beann
 \mathscr{FL}^*f^\nu_B &=& 
-\xi^\mu\derpar{^2\mathscr{L}}{x^\mu\partial y^B_\nu} - \xi^A\derpar{^2\mathscr{L}}{y^A\partial y^B_\nu}-\left(\derpar{\xi^A}{x^\mu}-y^A_\rho\derpar{\xi^\rho}{x^\mu}+y^C_\mu\frac{\partial \xi^A}{\partial y^C}\right)\frac{\partial^2\mathscr{L}}{\partial y^A_\mu\partial y^B_\nu} \ , \\
&=&
\derpar{}{y^B_\nu}\left[\xi^\mu\derpar{\mathscr{L}}{x^\mu}+\xi^A\derpar{\mathscr{L}}{y^A} + \left(\derpar{\xi^A}{x^\mu}-y^A_\rho\derpar{\xi^\rho}{x^\mu}+y^C_\mu\frac{\partial \xi^A}{\partial y^C}\right)\derpar{\mathscr{L}}{y^A_\mu}\right] \\ 
& &-\derpar{}{y^B_\nu}\left(\derpar{\xi^A}{x^\mu}-y^A_\rho\derpar{\xi^\rho}{x^\mu}+y^C_\mu\frac{\partial \xi^A}{\partial y^C}\right)\derpar{\mathscr{L}}{y^A_\mu} \ ,\\
&=&
\derpar{}{y^B_\nu}\left[\xi^\mu\derpar{\mathscr{L}}{x^\mu}+\xi^A\derpar{\mathscr{L}}{y^A} + \left(\derpar{\xi^A}{x^\mu}-y^A_\rho\derpar{\xi^\rho}{x^\mu}+y^C_\mu\frac{\partial \xi^A}{\partial y^C}\right)\derpar{\mathscr{L}}{y^A_\mu}\right] \\ 
& &-\left(\frac{\partial \xi^A}{\partial y^B}\derpar{\mathscr{L}}{y^A_\nu}-\derpar{\xi^\nu}{x^\mu}\derpar{\mathscr{L}}{y^B_\mu}\right) \ .
\eeann
Now recall that the variation of the Lagrangian density with respect to diffeomorphisms generated by $X_\xi$ is given by the Lie derivative,
\beann
\Lie(X_\xi){\bf L}&=& \Lie(X_\xi)(\mathscr{L}\,\d^mx)
\\ &=&
-\left[
\xi^\mu\derpar{\mathscr{L}}{x^\mu}+\xi^A\derpar{\mathscr{L}}{y^A}+\left(\derpar{\xi^A}{x^\mu}-y^A_\rho\derpar{\xi^\rho}{x^\mu}+y^C_\mu\derpar{\xi^A}{y^C}\right)\derpar{\mathscr{L}}{y^A_\mu}+\mathscr{L}\derpar{\xi^\mu}{x^\mu}\right]\d^mx.
\eeann
Then, if $\Lie(X_\xi){\bf L}=0$, it follows that 
$$
\mathscr{FL}^*f^\nu_B = - \derpar{}{y^B_\nu}\left(\mathscr{L}\derpar{\xi^\mu}{x^\mu}\right) -\frac{\partial \xi^A}{\partial y^B}\derpar{\mathscr{L}}{y^A_\nu}+\derpar{\xi^\nu}{x^\mu}\derpar{\mathscr{L}}{y^B_\mu} =
- \derpar{\xi^\mu}{x^\mu}\derpar{\mathscr{L}}{y^B_\nu}-\frac{\partial \xi^A}{\partial y^B}\derpar{\mathscr{L}}{y^A_\nu}+\derpar{\xi^\nu}{x^\mu}\derpar{\mathscr{L}}{y^B_\mu} \ ,
$$
hence,
\begin{equation*}
f^\nu_B = -\derpar{\xi^\mu}{x^\mu} p_B^\nu -\frac{\partial \xi^A}{\partial y^B}p_A^\nu+\derpar{\xi^\nu}{x^\mu}p_B^\mu \ .
\end{equation*}
Then with $f^\mu=-\xi^\mu$ and $f^A=-\xi^A$, the expression above is in agreement with the expression \eqref{liftJ1b} for the canonical lift $Y_\xi\in\vf(J^1\pi^*)$, and hence, $\mathscr{FL}_*X_\xi=Y_\xi$. 
\qed

Observe that, if $\Lie(X_\xi)\d^mx=0$, then 
$\Lie(X_\xi){\bf L}=0\, \Longleftrightarrow\,
\Lie(X_\xi)\mathscr{L}=0$.

\subsection{Lifting diffeomorphisms and vector fields to submanifolds of $J^1\pi^*$}
\label{lifting2}

Finally, we investigate the notion of canonical lifts to primary constraint submanifolds $P_\circ$ of $J^1\pi^*$ for
singular Hamiltonian field theories.
In general, it is not possible to canonically lift diffeomorphisms and vector fields from $E$ to any submanifold of $J^1\pi^*$
unless some additional condition is imposed.
Here, we present a preliminary definition for such lifts to $P_\circ$ under the rather restrictive condition that the lifted diffeomorphisms to $J^1\pi^*$ preserve the primary constraint submanifold.

As in Section \ref{MHFT},
let $\jmath_\circ\colon P_\circ\hookrightarrow J^1\pi^*$
be a $\tau$-transverse embedded submanifold of $J^1\pi^*$ so that
$P_\circ$ is a fibered submanifold over $E$ and $M$, 
with natural projections \eqref{projections}.
When such submanifolds arise from an almost-regular Lagrangian system, it follows that $P_\circ=\text{Im}{\mathscr F}\mathscr{L}$.

\begin{definition}
\label{liftdifeo0}
Let \, $\Phi_E\colon E\to E$ \, be a diffeomorphism which induces a diffeomorphism $\Phi_M\colon M\to M$, and let $\Phi_{J^1\pi^*}\colon J^1\pi^*\to J^1\pi^*$ be 
the canonical lift of $\Phi_E$ to $J^1\pi^*$.
If $\Phi_{J^1\pi^*}$ leaves $P_\circ$ invariant,
the \textbf{canonical lift of $\Phi_E$ to $P_\circ\subset J^1\pi^*$}
is the restriction of $\Phi_{J^1\pi^*}$ to $P_\circ$, 
$$
\Phi_{P_\circ}:=\Phi_{J^1\pi^*}\vert_{P_\circ}\colon P_\circ\to P_\circ\ .
$$
\end{definition}

The canonically lifted infinitesimal vector field generators of such diffeomorphisms is then defined as follows:

\begin{definition} 
\label{liftvf0}
Let $\xi_E\in\vf(E)$ be a $\pi$-projectable vector field 
whose flux consists of one-parameter groups of local diffeomorphisms of $E$, denoted $\phi_t$.
Furthermore, suppose that the canonical lift $Y_\xi\in\vf(J^1\pi^*)$ of $\xi_E$ to $J^1\pi^*$ is tangent to $P_\circ$.
Then, the \textbf{canonical lift of $\xi_E$ to $P_\circ\subset J^1\pi^*$} is the vector field $Y_\xi^\circ\in\vf(P_\circ)$ 
whose flux consists of the one-parameter groups of local diffeomorphisms $\phi_t^\circ$ which are the canonical lifts of $\phi_t$ to $P_\circ$,
and hence
$$(\jmath_\circ)_*Y_\xi^\circ=Y_\xi\vert_{P_\circ}\ .$$
\end{definition}

The geometrical setting is depicted in the following diagram:
\begin{equation}
    \begin{tikzcd}[column sep=1.75cm, row sep=1.25cm]
        \mathcal{M}\pi \arrow[ddr, swap, "\widetilde\sigma"] 
        \arrow[r, swap, "\sigma"]& J^1\pi^* \arrow[r, "\Phi_{J^1\pi^*}"] & J^1\pi^*  & \mathcal{M}\pi \arrow[ddl, "\widetilde\sigma"]
        \arrow[l, "\sigma"]\\
        & P_\circ \arrow[d, "\tau_\circ"] \arrow[r, "\Phi_{P_\circ}"] \arrow[u, swap, "\jmath_\circ"] & P_\circ \arrow[d, swap, "\tau_\circ"] \arrow[u, swap, "\jmath_\circ"] \\
        & E \arrow[r, swap, "\Phi_E"] & E 
    \end{tikzcd}
\end{equation}

Now, consider the particular case in which $(P_\circ,\Omega_{\mathscr{H}}^\circ)$ is
the Hamiltonian system associated to an almost-Lagrangian system
$(J^1\pi,\Omega_{\mathscr{L}})$
(as described in the last paragraph of Section \ref{MHFT})
with final constraint submanifold
$P_f\subseteq P_\circ$. 
In this case, if $S_f\subset J^1\pi$
is the Lagrangian final constraint submanifold,
then $\mathscr{FL}_\circ(S_f)=P_f$
(\cite{LMMMR-2005,GGR-2022,art:Roman09}).

A natural question to ask is if there exists a result analogous to Proposition \ref{propXYequivalence} for these kinds of singular field theories. 
However, in general, it is not the case that $\mathscr{FL}_{\circ*}X_\xi = Y^\circ_\xi$,
even when $\Lie(X_\xi){\bf L}=0$.
This is due to the fact that the canonical lift $X_\xi\in\vf(J^1\pi)$ is not always $\mathscr{FL}_\circ$-projectable on all of $J^1\pi$.
Moreover, if $\Lie(X_\xi){\bf L}=0$ and $X_\xi$ is $\mathscr{FL}_\circ$-projectable on all of $J^1\pi$, 
then we can ask if $\mathscr{FL}_{\circ *}X_\xi = Y^\circ_\xi$ always holds; 
this is difficult to prove in general, but the result holds at least in the singular field theories we study in this paper.

Fortunately, in some cases when $X_\xi$ is not $\mathscr{FL}_\circ$-projectable on all of $J^1\pi$, 
$X_\xi$ is at least $\mathscr{FL}_\circ$-projectable from the final Lagrangian constraint submanifold $S_f\subset J^1\pi$ which projects
onto the final Hamiltonian constraint submanifold $P_f\subset J^1\pi^*
$.
Moreover, as $\mathscr{FL}_\circ$ is a surjective submersion, 
if $X_\xi\in\vf(J^1\pi)$ is $\mathscr{FL}_\circ$-projectable and tangent to $S_f$,
then $(\mathscr{FL}_\circ)_*X_\xi\in\vf(P_\circ)$ is tangent to $P_f$.
Then, in such cases,
assuming that $\Lie(X_\xi){\bf L}=0$ and that $X_\xi$ is only $\mathscr{FL}_\circ$-projectable 
from points on $S_f\subset J^1\pi$, 
we ask if $\mathscr{FL}_{\circ*}X_\xi\vert_{S_f} = Y^\circ_\xi\vert_{P_f}$ always holds;
this is also difficult to prove in general, but is straightforward to check when working with a specific field theory.

In closing, if $\xi_E\in\vf(E)$ is a canonical lift
of a vector field $\xi_M\in\vf(M)$, then
the canonical lift of $\xi_E$ to ${\cal M}\pi$, $J^1\pi^*$,
and $P_\circ$ are also called the {\sl\textbf{canonical lifts}} of $\xi_M$ to ${\cal M}\pi$, $J^1\pi^*$,
and $P_\circ$, 
which are denoted $Z_\xi\in\vf({\cal M}\pi)$, $Y_\xi\in\vf(J^1\pi^*)$, 
and $Y_\xi^\circ\in\vf(P_\circ)$ respectively.
The same goes for canonical lifts of diffeomorphisms from $M$ to ${\cal M}\pi$, $J^1\pi^*$, and $P_\circ$.
Recall from the discussion at the beginning of Section \ref{lifting} that canonical lifts from $M$ to $E$ do not always exist.

\section{Hamiltonian symmetries}

In this section, we study symmetries along with their associated conserved quantities in the De Donder--Weyl Hamiltonian formalism for (pre)multisymplectic classical field theories.
We will be especially interested in symmetries which are natural lifts of base space diffeomorphisms
to the Hamiltonian phase space, the restricted multimomentum bundle.
For a more general introduction to this subject with a focus on the Lagrangian formalism for multisymplectic classical field theories, see \cite{GR-2023};
see also \cite{EMR-99b,art:GPR-2016} for other considerations.

\subsection{Hamiltonian Noether symmetries}

 \begin{definition}
 \label{Cartansym}
Let $(J^1\pi^*,\Omega_{\mathscr{H}})$ be a regular Hamiltonian system.
\ben
\item
A \textbf{Noether} or \textbf{Cartan symmetry}
of the regular Hamiltonian system $(J^1\pi^*,\Omega_{\mathscr{H}})$
is a diffeomorphism $\Phi\colon J^1\pi^*\to J^1\pi^*$
such that $\Phi^*\Omega_{\mathscr{H}}=\Omega_{\mathscr{H}}$.
In particular, when $\Phi^*\Theta_{\mathscr{H}}=\Theta_{\mathscr{H}}$,
the Hamiltonian Noether symmetry $\Phi$ is called
\textbf{exact}.

The Noether symmetry is said to be \textbf{natural} if $\Phi$ is the canonical lift of a diffeomorphism $\Phi_E\colon E\to E$
satisfying the condition of Definition \ref{liftdifeohat}.
\item
An \textbf{infinitesimal Noether} or \textbf{ Cartan symmetry} 
of a regular Hamiltonian system $(J^1\pi^*,\Omega_{\mathscr{H}})$
is a vector field $Y\in\vf(J^1\pi^*)$ for which $\Lie(Y)\Omega_{\mathscr{H}}=0$.
In particular, when $\Lie(Y)\Theta_{\mathscr{H}}=0$, 
the infinitesimal Hamiltonian Noether symmetry $Y$ is called 
\textbf{exact}.

The infinitesimal Noether symmetry is said to be \textbf{natural} if $Y$ is the canonical lift of a $\pi$-projectable vector field $\xi_E\in\vf(E)$.
\een
 \end{definition}

\begin{proposition}
Every (infinitesimal) natural Noether symmetry transforms  solutions of the Hamiltonian field equations into equivalent solutions.
\label{symsect2}
\end{proposition}
\proof
Let $\psi\colon M\to J^1\pi^*$
be a section solution to the field equations; that is, $\psi^*i(X)\Omega_{\mathscr{H}}=0$,
for every $X\in\vf(J^1\pi^*)$. 
If $\Phi\in{\rm Diff}(J^1\pi^*)$ is a natural Noether symmetry, then
it is the canonical lift of a diffeomorphism $\Phi_E\colon E\to E$
which restricts to a diffeomorphism $\Phi_M\colon M\to M$.
Therefore, $\Phi$ preserves the fibration $\overline{\tau}\colon J^1\pi^*\to M$ and, hence, $\Phi\circ\psi\circ\Phi_M^{-1}$
is a section of $\overline{\tau}$. Then,
\beann
(\Phi\circ\psi\circ\Phi_M^{-1})^*\inn(X)\Omega_{\mathscr{H}}&=&
(\Phi_M^{-1})^*[\psi^*(\Phi^*\inn(X)\Omega_{\mathscr{H}})]=
(\Phi_M^{-1})^*[\psi^*\inn(\Phi_*^{-1}X)(\Phi^*\Omega_{\mathscr{H}})]
\\ &=&
(\Phi_M^{-1})^*[\psi^*\inn(X')\Omega_{\mathscr{H}}]=0 \ ,
\eeann
since $\Phi_M^{-1}$ is a diffeomorphism and $\Phi^*\Omega_{\mathscr{H}}=\Omega_{\mathscr{H}}$, where $X'=\Phi_*^{-1}X\in\vf(J^1\pi^*)$ is an arbitrary vector field.
Therefore, $\Phi\circ\psi\circ\Phi_M^{-1}$ is also a section
solution to the field equations.

The proof for the infinitesimal case is a straightforward consequence of this result and the definition. 
\qed

\vspace{10pt}

Now, consider the case where $(P_\circ,\Omega_{\mathscr{H}}^\circ)$ is
a singular Hamiltonian system (see paragraph \ref{LagHam})
with final constraint submanifold $P_f \subseteq P_\circ$. 
Then, 

 \begin{definition}
 \label{Cartansym0}
 \ \
\ben
\item
A \textbf{Noether} or \textbf{Cartan symmetry}
of a singular Hamiltonian system $(P_\circ,\Omega^\circ_{\mathscr{H}})$
is a diffeomorphism $\Phi_\circ\colon P_\circ\to P_\circ$
such that $\Phi_\circ^*\Omega^\circ_{\mathscr{H}}=\Omega^\circ_{\mathscr{H}}$
(at least on $P_f$).
In particular, when $\Phi_\circ^*\Theta^\circ_{\mathscr{H}}=\Theta^\circ_{\mathscr{H}}$,
the Noether symmetry $\Phi_\circ$ is called
\textbf{exact}.

A Noether symmetry is said to be \textbf{natural} if $\Phi_\circ$ is the canonical lift to $P_\circ$ of a diffeomorphism $\Phi_E\colon E\to E$
satisfying the conditions of Definition \ref{liftdifeo0}.
\item
An \textbf{infinitesimal Noether} or \textbf{ Cartan symmetry} 
of a singular Hamiltonian system $(P_\circ,\Omega^\circ_{\mathscr{H}})$
is a vector field $Y^\circ\in\vf(P_\circ)$ for which $\Lie(Y^\circ)\Omega^\circ_{\mathscr{H}}=0$
(at least on $P_f$).
In particular, when $\Lie(Y^\circ)\Theta^\circ_{\mathscr{H}}=0$, 
the infinitesimal Hamiltonian Noether symmetry $Y^\circ$ is called 
\textbf{exact}.

An infinitesimal Noether symmetry is said to be \textbf{natural} if $Y^\circ$ is the canonical lift to $P_\circ$ of a $\pi$-projectable vector field $\xi_E\in\vf(E)$ satisfying the conditions of Definition \ref{liftvf0}.
\een
 \end{definition}

Geometric gauge symmetries are defined in singular Hamiltonian systems
analogously to how they are defined in singular Lagrangian systems \cite{GR-2023} as follows:
 
\begin{definition}
Let $(P_\circ,\Omega_\mathscr{H}^\circ)$ be a singular Hamiltonian system with final constraint submanifold $P_f\hookrightarrow P_\circ$.
A \textbf{geometric infinitesimal gauge symmetry} is a vector field $Y\in\vf(P_\circ)$ such that:
\ben
\item
$Y\in\ker\,\Omega_\mathscr{H}^\circ$; that is, $\inn(Y)\Omega_\mathscr{H}^\circ=0$.
\item
$Y$ is a $\overline{\tau}_\circ$-vertical vector field,  $Y\in\vf^{V(\overline{\tau}_\circ)}(P_\circ)$.
\item
$Y$ is tangent to $P_f$.
\een
\end{definition}

Local diffeomorphisms generated by the flow of Hamiltonian gauge vector fields 
are called {\sl \textbf{geometric gauge transformations}}.
Physical states (i.e., sections which are stable solutions to the Hamiltonian field equations) 
which are related to one another by such Hamiltonian geometric gauge transformations are called {\sl \textbf{geometric gauge equivalent states}}
and they are physically equivalent.

\subsection{Conserved quantities and multimomentum maps}

\begin{definition}
A \textbf{conserved quantity} of a regular Hamiltonian system $(J^1\pi^*,\Omega_{\mathscr{H}})$ is a form $\alpha\in\df^{m-1}(J^1\pi^*)$ which satisfies
$\Lie({\bf X}_{\mathscr{H}})\alpha=0$ (see Appendix \ref{append})
for every locally decomposable, $\overline{\tau}$-transverse multivector field ${\bf X}_{\mathscr{H}}\in\vf^m(J^1\pi^*)$ 
solution to the Hamiltonian field equations \eqref{equationHEOM}.
 \end{definition}
 
If $\alpha\in\df^{m-1}(J^1\pi^*)$ is a conserved quantity and ${\bf X}_{\mathscr{H}}\in\vf^m(J^1\pi^*)$
is a locally decomposable, $\overline{\tau}$-transverse multivector field solution to \eqref{equationHEOM},
then $\alpha$ is closed on the integral submanifolds of ${\bf X}_{\mathscr{H}}$; that is,
if $j_S\colon S\hookrightarrow J^1\pi^*$
 is an integral submanifold of ${\bf X}_{\mathscr{H}}$, then $\d j_S^*\alpha=0$.
Moreover, for every $\alpha\in\df^{m-1}(J^1\pi^*)$ and ${\bf X}_{\mathscr{H}}\in\vf^m(J^1\pi^*)$,
if $\psi\colon M\to J^1\pi^*$ is an  integral section of ${\bf X}_{\mathscr{H}}$ such that $\psi^*\alpha\in\df^{m-1}(M)$,
then there is a unique vector field $X_{\psi^*\alpha}\in\vf(M)$ such that
$\inn(X_{\psi^*\alpha})\omega=\psi^*\alpha$.
As the {\sl divergence} of $X_{\psi^*\alpha}$ is the function ${\rm div}X_{\psi^*\alpha}\in C^\infty(M)$
defined by $\Lie(X_{\psi^*\alpha})\omega= ({\rm div}X_{\psi^*\alpha})\,\omega$,
it follows that
$({\rm div}X_{\psi^*\alpha})\,\omega=\d{\psi^*\alpha}$.
Therefore, by the above property,  $\alpha$ is a conserved quantity if,  and only if, $\d{\psi^*\alpha}=0$ or, equivalently,
${\rm div}X_{\psi^*\alpha}=0$.
Then, on every bounded domain $U\subset M$,
{\sl Stokes theorem} leads to
$$
\int_{\partial U}{\psi^*\alpha}=\int_U \left({\rm div}X_{\psi^*\alpha}\right)\,\omega=\int_U\d{\psi^*\alpha}=0 \ .
$$
The form $\psi^*\alpha$ is called the {\sl \textbf{conserved current}} 
associated with the conserved quantity $\alpha$.

In this context, {\sl Noether's theorem} is stated as:

\begin{theorem}
{\rm Noether's Theorem:}
Let $(J^1\pi^*,\Omega_{\mathscr{H}})$ be a regular Hamiltonian system 
and $Y\in\vf (J^1\pi^*)$ an infinitesimal Hamiltonian Noether symmetry
with $\inn(Y)\Omega_{\mathscr{H}}=-\d\alpha_Y$ in an open set $U\subset J^1\pi^*$. 
Then,
for every locally decomposable, $\overline{\tau}$-transverse multivector field ${\bf X}_{\mathscr{H}}\in\vf^m(J^1\pi^*)$ 
solution to the field equations \eqref{equationHEOM}, it follows that
 $$
 \Lie({\bf X}_{\mathscr{H}})\alpha_Y=0 \ .
 $$
For every integral section $\psi$ of ${\bf X}_{\mathscr{H}}$,
 the form $\psi^*\alpha_Y$ is called a {\sl \textbf{Noether current}}.
 \label{Nth}
 \end{theorem}

Observe that, on $U$,
\beq
\Lie(Y)\Theta_{\mathscr{H}}=\d\inn(Y)\Theta_{\mathscr{H}}+\inn(Y)\d\Theta_{\mathscr{H}}=
\d\inn(Y)\Theta-\inn(Y)\Omega_{\mathscr{H}}=
\d\left(\inn(Y)\Theta_{\mathscr{H}}-\alpha_Y\right)\equiv\d\zeta_Y \ ,
 \label{closedTheta}
\eeq
and hence, if $Y$ is an exact infinitesimal Noether symmetry, 
then $\zeta_Y$ is closed and $\alpha_Y=-\inn(Y)\Theta_{\mathscr{H}}$.

We are mainly interested in Hamiltonian Noether symmetries which are
fiber-preserving diffeomorphisms of the bundle $J^1\pi^*\to M$
and in infinitesimal Hamiltonian Noether symmetries which are
$\overline{\tau}$-projectable vector fields.

It is usual that symmetries arise from the action $\Phi$ of a Lie group $G$ on the
regular Hamiltonian system $(J^1\pi^*,\Omega_{\mathscr{H}})$
which satisfies
$\Phi_g^*\Omega_{\mathscr{H}}=\Omega_{\mathscr{H}}$ for every $g\in G$, and
$\Lie(Y_{\xi})\Omega_{\mathscr{H}}= 0$ for every $\xi\in\mathfrak{g}$,
where $\mathfrak{g}$ is the Lie algebra of $G$ and $Y_\xi\in\vf(J^1\pi^*)$ is the vector field induced by $\xi\in\mathfrak{g}$.
Therefore, as $\Lie(Y_{\xi})\Omega_{\mathscr{H}}=\d\inn(Y_{\xi})\Omega_{\mathscr{H}}=0$,
it follows that, on an open neighbourhood $U\subset J^1\pi^*$ around every point in $J^1\pi^*$,
there exists a form ${\rm J}_\xi\in\df^{m-1}(U)$  such that, 
\begin{equation}
\label{equation:dJ}
\inn(Y_\xi)\Omega_{\mathscr{H}}=-\text{d}{\rm J}_\xi\ .
\end{equation}
The form ${\rm J}_\xi$ coincides with the conserved quantity $\alpha_Y$ given in the Noether theorem \ref{Nth} and is determined up to some exact form $\text{d}\beta_{\xi}$, with $\beta_{\xi}\in\df^{m-2}(U)$. 
Furthermore,
\begin{equation}
\label{equation:dJcartan}
\text{d}{\rm J}_{\xi}=-\inn(X_{\xi})\Omega_{\mathscr{H}}=\inn(X_{\xi})\text{d}\Theta_{\mathscr{H}}=\Lie(X_{\xi})\Theta_{\mathscr{H}}-\text{d}\inn(X_{\xi})\Theta_{\mathscr{H}} \ ,
\end{equation}
and, as $Y_{\xi}$ is an infinitesimal Noether symmetry,
\eqref{closedTheta} holds for $Y_{\xi}$:
$\Lie(Y_{\xi})\Theta_{\mathscr{H}}=\text{d}\alpha_{\xi}$ for some $\alpha_{\xi} \in \df^{m-1}(U)$. 
It thereby follows that, in general, 
$$
{\rm J}_{\xi}=-\inn(X_{\xi})\Theta_{\mathscr{H}}+\alpha_{\xi}+\text{d}\beta_{\xi} \ .
$$

\begin{definition}
For every point ${\rm p}\in J^1\pi^*$, define the linear map
$$
\begin{array}{ccccc}
\left.{\rm J}\right\vert_{\rm p}&\colon&{\bf g}^*&\to&\df^{m-1}(\Tan_{\rm p}J^1\pi^*)
\\
& &\xi&\mapsto&{\rm J}\vert_{\rm p}(\xi):={\rm j}_{\xi}({\rm p})\ .
\end{array} 
$$
Then, the map 
$$
\begin{array}{ccccc}
{\rm J}&\colon& J^1\pi^*&\to&{\bf g}^*\otimes\df^{m-1}(J^1\pi^*)
\\
& & {\rm p}&\mapsto & {\rm J}({\rm p}):={\rm J}\vert_{\rm p}
\end{array}
$$
is called the \textbf{multimomentum map}
associated with the symmetry group $G$.
\end{definition}

The terminology ``multimomentum map'' is also used to refer the $(m-1)$-form ${\rm J}_\xi$ given in \eqref{equation:dJ}. 
Furthermore, it is usual to specify the multimomentum map by
${\rm J}_{\xi}({\rm p})=\left<{\rm J}({\rm p}),\xi\right>$.
Thus, Noether symmetries on $J^1\pi^*$ produce multimomentum maps on $J^1\pi^*$
and the so-called {\sl \textbf{Noether current}} ${\rm j}=j^\mu\text{d}^{m-1}x_\mu\in \df^{m-1}(M)$ of a symmetry 
is specified as ${\rm j}=\psi^*{\rm J}_\xi$.

For singular Hamiltonian systems $(P_\circ,\Omega_{\mathscr{H}}^\circ)$
all these concepts are defined analogously on $P_\circ$.

\section{Examples of Field Theories}

In the following sections, we give some examples of classical field theories which exhibit symmetries under the lifted action of base space diffeomorphisms.

\subsection{The Klein--Gordon Field}

In this section, we study the multisymplectic formulation for a real scalar field on a $m$-dimensional spacetime manifold $M=\mathbb{R}^m$, equipped with the Minkowski metric.
The configuration manifold $E=M\times\mathbb{R}$ is equipped with projection map $\pi:E \rightarrow M$. 
The local coordinates on $M$ are $x^\mu$ and $E$ comes with natural coordinates $(x^\mu, \varphi)$.
Recall that the Minkowski metric $\eta_{\mu\nu}$ has signature $(-+...+)$ and the spacetime partial derivatives are also denoted as $\partial_\mu = \partial/\partial x^\mu$.

\subsubsection{Lagrangian Formalism}

The Lagrangian formalism \cite{EMR-96,Gc-73,book:Saunders89} takes place on the first-order jet bundle $J^1\pi$ which has natural coordinates  $(x^\mu, \varphi, \varphi_\mu)$. 
The Lagrangian density $\textbf{L}$ and Lagrangian function $\mathscr{L}(x^{\mu},\varphi, \varphi_{\mu})$ on $J^1\pi$ for the scalar field theory are
\begin{equation}
\textbf{L}=\mathscr{L}(x^{\mu},\varphi, \varphi_{\mu})\text{d}^mx=-\frac{1}{2}\left(\eta^{\mu\nu}\varphi_\mu \varphi_\nu + m^2\varphi^2\right) \text{d}^mx\ ,
\end{equation}
from which the Lagrangian energy $E_\mathscr{L}$ is given by
\begin{equation}
E_{\mathscr{L}}\equiv \frac{\partial \mathscr{L}}{\partial \varphi_{\mu}} \varphi_{\mu} - \mathscr{L}=-\frac{1}{2}\eta^{\mu\nu}\varphi_\mu \varphi_\nu + \frac{1}{2}m^2\varphi^2\ .
\end{equation}
This is a regular Lagrangian system in the sense that the Hessian
\begin{equation}
\label{eq:phihessian}
\frac{\partial^2\mathscr{L}}{\partial\varphi_\mu\partial\varphi_\nu}= -
 \eta^{\mu\nu}
\end{equation}
is regular. 
The jet bundle $J^1\pi$ comes equipped with the Poincar\'e-Cartan $m$-form $\Theta_\mathscr{L}$ and the multisymplectic $(m+1)$-form $\Omega_\mathscr{L}$
which are given by, 
\beann
\Theta_\mathscr{L} &=& \eta^{\mu\nu}\left(\frac{1}{2}\varphi_\mu\varphi_\nu\text{d}^mx-\varphi_\nu\text{d}\varphi\wedge\text{d}^{m-1}x_\mu\right)\ ,\\
\Omega_{\mathscr{L}}&=& \eta^{\mu\nu} \text{d}\varphi_\nu\wedge \left( \text{d}\varphi\wedge \text{d}^{m-1}x_\mu - \varphi_\mu\text{d}^mx\right) \ .
\eeann
The multisymplectic form $\Omega_\mathscr{L}$ is $1$-non-degenerate as a result of the regularity of the Hessian \eqref{eq:phihessian}.

The field equations are produced by a class of multivector fields $\{{\bf X}_{\mathscr{L}}\} \subset \mathfrak{X}^m(J^1\pi)$ 
that are locally decomposable, integrable, and $\bar{\pi}^1-$transverse $\left(\inn({\bf X}_\mathscr{L})\text{d}^mx\neq 0\right)$.
We use a representative of $\{{\bf X}_{\mathscr{L}}\}$ for which $\inn({\bf X}_{\mathscr{L}})\text{d}^mx=1$,
expressed locally as
\begin{equation}
\label{Xlag}
{\bf X}_\mathscr{L}=\bigwedge^{m-1}_{\mu=0} X_\mu = \bigwedge^{m-1}_{\mu=0}\left( \frac{\partial}{\partial x^\mu} + D_\mu \frac{\partial}{\partial \varphi} + H_{\mu\nu}\frac{\partial}{\partial \varphi_\nu}\right)\ .
\end{equation}
The field equations are then given as $\inn({\bf X}_{\mathscr{L}})\Omega_{\mathscr{L}}=0$,
from which it follows that
\begin{subequations}
\begin{align}
\eta^{\mu\nu}H_{\mu\nu} - m^2\varphi=0\ , \label{eq:phiLeqsa}    \\
\eta^{\mu\nu}\left(D_\mu-\varphi_\mu\right)=0\ .  
\label{eq:phiLeqsb} 
\end{align}
\end{subequations}
Furthermore, the solution to the variational problem on $J^1\pi$ requires that $X_{\mathscr{L}}$ must be {\sl holonomic} multivector fields
(the {\sc sopde} condition).
By definition, this means that they have holonomic integral sections 
$j^1\phi:M\rightarrow M\times\mathds{R}:x^\mu\mapsto(x^\mu, \varphi, \partial_\mu \varphi)$,
so $D_\mu=\partial_\mu\varphi$. 
Therefore, in the expression \eqref{Xlag} must be $D_\mu=\varphi_\mu$,
and this condition shows up in the field equations \eqref{eq:phiLeqsb}; 
this is expected as the {\sc sopde} condition always shows up in the field equations when the Lagrangian is regular. 
The final field equations are obtained from \eqref{eq:phiLeqsa} on holonomic integral sections, giving:
\begin{equation}
\label{eq:curvedkg}
\eta^{\mu\nu}\partial_\mu\partial_\nu\varphi - m^2\varphi=0\ .
\end{equation}

\subsubsection{Hamiltonian Formalism}

The De Donder--Weyl Hamiltonian formulation of Klein--Gordon field theory is constructed on the extended multimomentum bundle ${\cal M}\pi$ which has local coordinates $\left(x^\mu,\varphi,p_\mu, p\right)$ and the restricted multimomentum bundle $J^1\pi^*$
which has natural coordinates $(x^\mu,\varphi,p_\mu)$.
The Klein--Gordon field theory is specified by a Hamiltonian section ${\rm h}=\big(x^\mu,\varphi,p_\mu, p= \frac{1}{2}\eta_{\mu\nu}p^\mu p^\nu - \frac{1}{2}m^2\varphi^2\big)$ so that the De Donder--Weyl Hamiltonian is given as
\begin{equation}
\mathscr{H}= - \frac{1}{2}\eta_{\mu\nu}p^\mu p^\nu + \frac{1}{2}m^2\varphi^2 \in C^\infty(J^1\pi^*)\ .
\end{equation}
Furthermore, this Hamiltonian field theory is related to its Lagrangian version by the 
following Legendre map:
\begin{equation*}
\mathscr{FL}^*p^{\mu}= \frac{\partial\mathscr{L}}{\partial \varphi_{\mu}}=- \eta^{\mu\nu}\varphi_\nu\ ,
\end{equation*}
which is invertible since the Hessian \eqref{eq:phihessian} is regular
($\mathscr{FL}$ is a local diffeomorphism); it follows that
\begin{equation}
\left(\mathscr{FL}^{-1}\right)^*\varphi_{\mu}=-\eta_{\mu\nu}p^\nu\ .
\end{equation}
Furthermore, the De Donder--Weyl Hamiltonian can be produced by the Legendre map as,
\begin{equation}
\mathscr{H} = p^{\mu}\left(\mathscr{FL}^{-1}\right)^*\varphi_{\mu}-\left(\mathscr{FL}^{-1}\right)^*\mathscr{L}= - \frac{1}{2}\eta_{\mu\nu}p^\mu p^\nu + \frac{1}{2}m^2\varphi^2 \ .
\end{equation}

The bundle $J^1\pi^*$ is equipped with the Hamilton--Cartan $m$-form $\Theta_\mathscr{H}$ and multisymplectic $(m+1)$-form $\Omega_{\mathscr{H}}$ given by
\beann
\Theta_\mathscr{H}&=&p^\mu\text{d}\varphi\wedge\text{d}^{m-1}x_\mu+\frac{1}{2}\eta_{\mu\nu}p^\mu p^\nu\text{d}^mx\ ,\\
\Omega_{\mathscr{H}}&=&-\text{d}p^\mu\wedge\text{d}\varphi\wedge\text{d}^{m-1}x_\mu - \left(\eta_{\mu\nu}p^\nu\text{d}p^\mu - m^2\varphi \text{d}\varphi \right)\wedge\text{d}^mx\ .
\eeann
The field equations are again produced by a class of multivector fields $\{{\bf X}_\mathscr{H}\}\subset \mathfrak{X}^m(J^1\pi^*)$ 
that are locally decomposable, integrable, and $\bar{\tau}$-transverse 
which satisfy $i({\bf X}_{\mathscr{H}})\Omega_{\mathscr{H}}=0$ for every ${\bf X}_{\mathscr{H}} \in \{{\bf X}_{\mathscr{H}}\}$. 
As above, we use a representative of $\{{\bf X}_{\mathscr{H}}\}$,
expressed as,
\begin{equation}
{\bf X}_\mathscr{H} = \bigwedge^{m-1}_{\mu=0}X_\mu = \bigwedge^{m-1}_{\mu=0} \left(\frac{\partial}{\partial x^\mu} + D_\mu \frac{\partial}{\partial \varphi} + H_{\mu}^\rho\frac{\partial}{\partial p^\rho}\right)\ ,
\end{equation}
and the field equations $i({\bf X}_{\mathscr{H}})\Omega_{\mathscr{H}}=0$ give
\beann
&H_\mu^\mu+ m^2\varphi=0\ ,  \\ 
&D_\mu+ \eta_{\mu\nu}p^\nu=0\ .
\eeann
Recall that the solution to the variational problem in the De Donder--Weyl Hamiltonian formalism requires that $\psi$ are integral sections of ${\bf X}_{\mathscr{H}}$, hence $D_\mu=\partial_\mu\varphi$ and $H_\mu^\rho=\partial_\mu p^\rho$,
giving the Hamilton--De Donder--Weyl equations:
\beann
&\partial_\mu p^\mu + m^2\varphi=0\ ,  \\
&\partial_\mu\varphi+ \eta_{\mu\nu}p^\nu=0\ . \\
\eeann

\subsubsection{Symmetries}

The symmetry group of the Klein--Gordon field theory on Minkowski spacetime consists of Lorentz transformations, written as
$$
x^\mu \rightarrow x^\mu - \Lambda^\mu_{\ \nu}x^\nu \quad , \quad \varphi_\mu(x) \rightarrow \varphi_\mu(x) + \Lambda^\nu_{\ \mu}\varphi_\nu(x) \ ,
$$
where $\Lambda^\mu_{\ \nu}$ are the infinitesimal Lorentz matrices which are antisymmetric. 
It follows that the vector fields which generate the Lorentz transformations on spacetime $M$ and the configuration bundle $E$ are,
$$
\xi = - \Lambda^\mu_{\ \nu}x^\nu\frac{\partial}{\partial x^\mu} \quad , \quad \xi_E = - \Lambda^\mu_{\ \nu}x^\nu\frac{\partial}{\partial x^\mu} \ ,
$$
respectively. 
The canonical lifts $X_\xi\in \mathfrak{X}(J^1\pi)$, $Z_\xi\in\vf(\mathcal {M}\pi)$, and $Y_\xi\in \mathfrak{X}(J^1\pi^*)$ are given by \eqref{j1Y}, \eqref{liftJ1}, and \eqref{liftJ1b}, respectively, and they are
\beann
X_\xi&=&- \Lambda^\mu_{\ \nu}x^\nu\frac{\partial}{\partial x^\mu} +  \Lambda^\nu_{\ \mu} \varphi_\nu\frac{\partial}{\partial\varphi_\mu}\ , \\
Z_\xi&=& -\Lambda^\mu_{\ \nu}x^\nu\frac{\partial}{\partial x^\mu}- \Lambda^\mu_{\ \nu} p^\nu
\frac{\partial}{\partial p^\mu}\ , \\
Y_\xi&=& -\Lambda^\mu_{\ \nu}x^\nu\frac{\partial}{\partial x^\mu}- \Lambda^\mu_{\ \nu} p^\nu
\frac{\partial}{\partial p^\mu}\ .
\eeann
Furthermore, $\mathscr{FL}_*X_\xi=Y_\xi$ trivially
(see also Proposition \ref{FLproj}).

It is straightforward to check that 
$\Lie(X_\xi)\textbf{L} = 0$,
and hence $\Lie(X_\xi)\Theta_\mathscr{L}=0$.
Therefore, $\Lie(Y_\xi)\Theta_\mathscr{H}=0$;
that is, both $X_\xi$ and $Y_\xi$ are natural exact Lagrangian and Hamiltonian infinitesimal Noether symmetries, respectively. 
The corresponding momentum maps are: 
\begin{align}
J_\mathscr{L}(X_\xi)&=-i(X_\xi)\Theta_\mathscr{L}= \Lambda^\rho_{\ \sigma}x^\sigma\eta^{\mu\nu}\left(\frac{1}{2}\varphi_\mu\varphi_\nu\text{d}^{m-1}x_\rho + \varphi_\nu\text{d}\varphi\wedge\text{d}^{m-2}x_{\mu\rho}\right) \in \Omega^{m-1}(J^1\pi)\ ,
\\
J_\mathscr{H}(Y_\xi)&=-i(Y_\xi)\Theta_\mathscr{H}
=\Lambda^\rho_{\ \sigma}x^\sigma\left( p^\mu\text{d}\varphi\wedge\text{d}^{m-2}x_{\mu\rho}+\frac{1}{2}\eta_{\mu\nu}p^\mu p^\nu\text{d}^{m-1}x_\rho \right)\in \Omega^{m-1}(J^1\pi^*)\ .
\end{align}

\subsection{The Polyakov String Action}

In the following section, the premultisymplectic treatment of the relativistic string on a Minkowski background spacetime is carried out starting from the famous {\sl Polyakov action}.
The premultisymplectic constraint analysis of this theory has been carried out previously \cite{LMD-2002}, using Ehresmann connections to write the field equations intrinsically. 
Here, we carry out the premultisymplectic constraint analysis using multivector fields.

\subsubsection{Lagrangian Formalism}

Consider the string worldsheet $\Sigma$ with coordinates $\sigma^a$ $(a=0,1)$ and auxiliary metric $g_{ab}$; partial worldsheet derivatives are denoted as $\partial_a=\partial/\partial\sigma^a$.
The embedding of the string worldsheet in spacetime $M$ (which has coordinates $x^\mu$) is specified by local maps $\Sigma \rightarrow M$ which are denoted as $x^\mu = x^\mu(\sigma)$. 
The variational fields considered here are the embedding maps $x^\mu(\sigma)$ and the components of the auxiliary inverse metric $g^{ab}(\sigma)$. 
It follows that the configuration bundle is specified as $\pi: E \rightarrow \Sigma: (\sigma^a, x^\mu, g^{ab})\mapsto \sigma^a$, and its first-order jet bundle $J^1\pi$ has local coordinates $(\sigma^a, x^\mu, g^{ab}, x^\mu_a, g^{ab}_c)$.

The {\sl Polyakov Lagrangian} function is written as
\begin{equation}
\label{PolyakovLag}
\mathscr{L}=-\frac{T}{2}\sqrt{-g\,}\eta_{\mu\nu}g^{ab}x^\mu_a x^\nu_b\in C^\infty(J^1\pi)\ .
\end{equation}
The resulting Lagrangian energy is 
\begin{equation}
E_\mathscr{L}=-\frac{T}{2}\sqrt{-g\,}\eta_{\mu\nu}g^{ab}x^\mu_a x^\nu_b\in C^\infty(J^1\pi)\ ,
\end{equation}
and the Poincar\'e--Cartan forms are:
\beann
\Theta_\mathscr{L}&=&-\frac{T}{2}\sqrt{-g\,}\eta_{\mu\nu}g^{ab}x^\nu_b\text{d}x^\mu\wedge\text{d}^1\sigma_a+\frac{T}{2}\sqrt{-g\,}\eta_{\mu\nu}g^{ab}x^\mu_a x^\nu_b\text{d}^2\sigma\ ,\\
\Omega_\mathscr{L}&=&T\sqrt{-g\,}\eta_{\mu\nu}\left(g^{ab}\text{d}x^\nu_b\wedge\text{d}x^\mu+x^\nu_b\text{d}g^{ab}\wedge\text{d}x^\mu 
- \frac{1}{2}g^{ab}x^\nu_b g_{dc}\text{d}g^{cd}\wedge\text{d}x^\mu
\right)\wedge\text{d}^1\sigma_a \\
& &-\frac{T}{2}\sqrt{-g\,}\eta_{\mu\nu}\left(x^\mu_a x^\nu_b\text{d}g^{ab}+2g^{ab}x^\mu_a \text{d}x^\nu_b
- \frac{1}{2}g^{ab}x^\mu_a x^\nu_b g_{dc}\text{d}g^{cd}
\right)\wedge\text{d}^2\sigma\ .
\eeann
The multi-Hessian for the Polyakov Lagrangian \eqref{PolyakovLag} is given by,
\begin{equation}
\label{PolyakovHessian}
\derpar{^2\mathscr{L}}{x^\mu_a\partial x^\nu_b}=-T\sqrt{-g\,}\eta_{\mu\nu}g^{ab}\quad , \quad \derpar{^2\mathscr{L}}{g^{ab}_i\partial g^{cd}_j}= \derpar{^2\mathscr{L}}{x^\mu_a\partial g^{cd}_j}=0\ ,
\end{equation}
which is clearly singular.

Now, we use a $2$-multivector field, 
\begin{equation}
{\bf X}_\mathscr{L}= \bigwedge^{1}_{a=0} X_a = \bigwedge^{1}_{a=0}\left(\derpar{}{\sigma^a}+A^\mu_a\derpar{}{x^\mu}+G^{bc}_a\derpar{}{g^{bc}}+D^\nu_{ab}\derpar{}{x^\nu_b}+K^{bc}_{ad}\derpar{}{g^{bc}_d}\right)\in\vf^2(J^1\pi)\ ,
\end{equation}
to calculate $i({\bf X}_\mathscr{L})\Omega_\mathscr{L}=0$, and obtain
\begin{align}
\label{stringeq1}
\begin{split}
0&=\sqrt{-g\,}\eta_{\mu\nu}g^{ab}(x^\mu_a-A^\mu_a)\ ,
\end{split}
\\
\begin{split}
\label{stringeq2}
0&=\sqrt{-g\,}\eta_{\mu\nu}\left(x^\mu_a x^\nu_b-2x^\nu_bA^\mu_a-g^{cd}g_{ab}x^\nu_dA^\mu_c - \frac{1}{2}g^{cd}g_{ba}x^\mu_cx^\nu_d\right)\ ,
\end{split}
\\
\begin{split}
\label{stringeq3}
0&=\sqrt{-g\,}\Big(\eta_{\mu\nu}g^{ab}D^\nu_{ab}+\eta_{\mu\nu}x^\nu_bG^{ab}_a - \frac{1}{2}\eta_{\mu\nu}g^{ab}g_{cd}x^\nu_bG^{cd}_a\Big)\ .
\end{split}
\end{align}
Equations \eqref{stringeq1} lead to one of the {\sc sopde} conditions,
$A^\mu_a=x^\mu_a$,
and they need not be imposed.
However, the remaining {\sc sopde} condition, $G^{bc}_a=g^{bc}_a$, must be imposed {\sl ad hoc}.
It is evident that imposing this last {\sc sopde} condition only affects equation \eqref{stringeq3}.
Upon combining equations \eqref{stringeq1} and \eqref{stringeq2} and imposing the aforementioned {\sc sopde} condition, $G^{bc}_a=g^{bc}_a$, the equations become:
\begin{align}
\begin{split}
0&=\sqrt{-g\,}\eta_{\mu\nu}\left(x^\mu_a x^\nu_b - \frac{1}{2}g^{cd}g_{ba}x^\mu_cx^\nu_d\right)\ ,\label{compatL1}
\end{split}
\\
\begin{split}
0&=\sqrt{-g\,}\Big(\eta_{\mu\nu}g^{ab}D^\nu_{ab}+\eta_{\mu\nu}x^\nu_b g^{ab}_a - \frac{1}{2}\eta_{\mu\nu}g^{ab}g_{cd}x^\nu_b g^{cd}_a\Big)\ .
\end{split}
\label{sopde3}
\end{align}
No {\sc sopde} constraints arise from \eqref{sopde3}. 
However, equation \eqref{compatL1}, which arises from combining equations \eqref{stringeq1} and \eqref{stringeq2}, is a compatibility constraint.
Similarly to Einstein--Cartan gravity, no stability (tangency) constraints arise from demanding that the Lie derivative of the constraint function \eqref{compatL1} with respect to the vector fields $X_a$ of the multivector field ${\bf X}_\mathscr{L}$ vanish on the submanifold defined by \eqref{compatL1}.
It follows that the compatibility constraint submanifold is the final constraint submanifold $S_f\hookrightarrow J^1\pi^*$ of this Lagrangian system. 
The final field equations on integral sections of ${\bf X}_\mathscr{L}$ are written as,
\begin{align}
\begin{split}
0&=\sqrt{-g\,}\eta_{\mu\nu}\left(\partial_a x^\mu\partial_b x^\nu-\frac{1}{2}g^{cd}g_{ba}\partial_c x^\mu\partial_d x^\nu\right)\ ,
\label{stringEL1}
\end{split}
\\
\begin{split}
0&=\sqrt{-g\,}\left(\eta_{\mu\nu}g^{ab}\partial_a\partial_b x^\nu+\eta_{\mu\nu}\partial_b x^\nu\partial_ag^{ab}-\frac{1}{2}\eta_{\mu\nu}g^{ab}g_{cd}\partial_b x^\nu\partial_a g^{cd}\right)\ .
\label{stringEL2}
\end{split}
\end{align}

Equation \eqref{stringEL2} is the Euler--Lagrange equation for $x^\mu$ 
while  \eqref{stringEL1} is the Euler--Lagrange equation for $g^{ab}$ which is well-known in the physics literature as the {\sl Virasoro constraint} (see e.g. \cite{BBS}).

It should be noted that the {\sl Hilbert stress--energy tensor} on $J^1\pi$ is given as
\begin{equation}
\label{StringSE}
T_{ab} = - \frac{2}{T}\frac{1}{\sqrt{-g}}\derpar{\mathscr{L}}{g^{ab}} = \eta_{\mu\nu} \left( x^\mu_a x^\nu_b - \frac{1}{2}g^{cd}g_{ba}x^\mu_cx^\nu_d \right) \ .
\end{equation}
The vanishing of this stress--energy tensor gives the compatibility constraint \eqref{compatL1}.
Notice that $T_{ab}$ is traceless everywhere on $J^1\pi$ and not just on the compatibility constraint as expected. 
In conclusion, composing the compatibility constraint \eqref{compatL1} with sections $j^1\phi:\Sigma\to J^1\pi$ gives the well-known Virasoro constraint and composing the 
{\sl Hilbert stress--energy tensor} \eqref {StringSE} with sections $j^1\phi$ gives the stress-energy tensor that is well-known in the string theory literature \cite{BBS};
this is in accordance with the reoccurring theme that we work on the premultisymplectic fiber bundle $J^1\pi$ 
while in the physics literature, field theory is carried out on the space of sections of $J^1\pi$ over the base space (which in this case is the string worldsheet $\Sigma$).

\subsubsection{Hamiltonian Formalism}

Since the Polyakov Lagrangian is singular, the Hamilton--De Donder--Weyl formalism takes place on a primary constraint submanifold $P_\circ\hookrightarrow J^1\pi^*$. 
The local coordinates on ${\cal M}\pi$ and $J^1\pi^*$ are $(\sigma^a, x^\mu, p^a_\mu, \pi^c_{ab}, p)$ and $(\sigma^a, x^\mu, p^a_\mu, \pi^c_{ab})$ respectively. 
The Legendre map gives,
\begin{equation}
\label{StringLT}
\mathscr{FL}^*p^a_\mu=\derpar{\mathscr{L}}{x^\mu_a}=-T\sqrt{-g\,}\eta_{\mu\nu}g^{ab} x^\nu_b \quad , \quad \mathscr{FL}^*\pi^c_{ab}=\derpar{\mathscr{L}}{g^{ab}_c}=0\ ,
\end{equation}
for the multimomenta.
The first equation above can be inverted to solve for the multivelocities in terms of the multimomenta.
The second equation above specifies the primary constraint submanifold $P_\circ\hookrightarrow J^1\pi^*$.

The De Donder--Weyl Hamiltonian $\mathscr{H}_\circ\in C^\infty(P_\circ)$ which satisfies $E_\mathscr{L}=\mathscr{FL}^*_\circ\mathscr{H}_\circ$ is given as, 
\begin{equation}
\mathscr{H}_\circ=-\frac{1}{2T\sqrt{-g}}\eta^{\mu\nu}g_{ab}p^a_\mu p^b_\nu\ ,
\end{equation}
and the corresponding Hamiltonian section is given as 
$h_\circ = (\sigma^a, x^\mu, p^a_\mu, \pi^c_{ab}, p=-\mathscr{H}_\circ)$.

The corresponding Hamilton--Cartan forms are, 
\beann
\Theta^\circ_\mathscr{H}&=&p^a_\mu\d x^\mu\wedge\d^{1}\sigma_a+\frac{1}{2T\sqrt{-g}}\eta^{\mu\nu}g_{ab}p^a_\mu p^b_\nu\d^2\sigma\ , \\
\Omega^\circ_\mathscr{H}&=&-\d p^a_\mu\wedge\d x^\mu\wedge\d^{1}\sigma_a  - \frac{1}{2T\sqrt{-g}}\eta^{\mu\nu} \left[2g_{ab} p^b_\nu\d p^a_\mu + \left(p^a_\mu p^b_\nu - \frac{1}{2}p^d_\mu p^c_\nu g_{cd}g^{ba}\right)\d g_{ab}\right]\wedge\d^2\sigma\ .
\eeann
Now use a $2$-multivector field,
\begin{equation}
{\bf X}^\circ_\mathscr{H}=\bigwedge^{1}_{a=0} X_a= \bigwedge^{1}_{a=0} \left(\derpar{}{\sigma^a}+A^\mu_a\derpar{}{x^\mu}+G^{bc}_a\derpar{}{g^{bc}}+D^b_{a \nu }\derpar{}{p_\nu^b}\right)\in\vf^2(P_\circ)\ ,
\end{equation}
to calculate $\inn({\bf X}^\circ_\mathscr{H})\Omega^\circ_\mathscr{H}=0$; then,
$$
0=A^\mu_a+\frac{1}{T\sqrt{-g\,}}\eta^{\mu\nu} g_{ab}p^b_\nu \quad , \quad
0= D^a_{a\mu} \quad , \quad
0= \frac{1}{2T\sqrt{-g}}\eta^{\mu\nu}\left(p^a_\mu p^b_\nu - \frac{1}{2}g_{cd}g^{ba}p^d_\mu p^c_\nu \right)\ .
$$
The third equation above is the Hamiltonian compatibility constraint which is 
the $\mathscr{FL}$ projection of the Lagrangian
compatibility constraint \eqref{compatL1}.
As in the Lagrangian formalism, no (tangency) constraints arise since the tangency conditions (i.e., the Lie derivatives of this compatibility constraint function with respect to the vector fields $X_a$) provide equations for the coefficients $D^b_{a \nu }$.
The third equation thereby defines the final constraint submanifold $P_f\hookrightarrow P_\circ \hookrightarrow J^1\pi^*$.  
Working on integral sections of the multivector field ${\bf X}^\circ_\mathscr{H}$ gives the Hamilton--De Donder--Weyl equations:
$$
0=\partial_a x^\mu+\frac{1}{T\sqrt{-g\,}}\eta^{\mu\nu} g_{ab}p^b_\nu \quad , \quad
0= \partial_a p^a_\mu \quad , \quad
0= \frac{1}{2T\sqrt{-g}}\eta^{\mu\nu}\left(p^a_\mu p^b_\nu - \frac{1}{2}g_{cd}g^{ba}p^d_\mu p^c_\nu \right)\ .
$$
The first equation gives the expression for the multivelocities written in terms of the multimomenta; this equation is the inverse of the equation obtained from the Legendre transform.
The second equation above is equivalent to the Euler--Lagrange equation \eqref{stringEL2} in the Lagrangian formalism. 
And finally, the third equation is the De Donder--Weyl version of the Virasoro constraint \eqref{stringEL1}.

\subsubsection{Symmetries}

The Polyakov string action exhibits three different symmetries: {\sl worldsheet diffeomorphisms, spacetime Poincar\'e transformations,} and {\sl Weyl transformations}. 
Each of these transformations generate exact premultisymplectomorphisms.

{\sl Worldsheet diffeomorphisms\/}:

The worldsheet diffeomorphisms are produced by a general change of coordinates on the string worldsheet as,
\begin{equation}
\label{eq:stringdiff}
\tilde{\sigma}^a =\sigma^a+\xi^a\ \Longrightarrow\ \partial_b \tilde{\sigma}^a=\delta^a_b+\partial_b\xi^a \ .
\end{equation}
The fields $x^\mu(\sigma)$ and $g^{ab}(\sigma)$ transform under the worldsheet diffeomorphisms as,
\begin{equation*}
x^\mu\rightarrow x^\mu - \xi^c\partial_c x^\mu \quad , \quad g^{ab}\rightarrow g^{ab} -\xi^c\partial_c g^{ab} + g^{ac}\partial_c \xi^b + g^{cb}\partial_c \xi^a \ .
\end{equation*}
The worldsheet diffeomorphisms are generated by the vector field $\xi_\Sigma\in\vf(\Sigma)$ given by
\begin{equation*}
\xi_\Sigma=-\xi^a\derpar{}{\sigma^a}\in\vf(\Sigma)\ ,
\end{equation*}
whose canonical lift to the configuration bundle $E$ is given by
\begin{equation*}
\xi_E=-\xi^a\frac{\partial}{\partial\sigma^a} - \left(g^{ac}\partial_c \xi^b + g^{cb}\partial_c \xi^a\right)\derpar{}{g^{ab}} \in\mathfrak{X}(E)\ .
\end{equation*}
The canonical lift $X_\xi\in \mathfrak{X}(J^1\pi)$ of $\xi_E$ to $J^1\pi$, given by \eqref{j1Y}, is
\begin{equation*}
X_\xi=-\xi^a\frac{\partial}{\partial\sigma^a} - \left(g^{ac}\partial_c \xi^b + g^{cb}\partial_c \xi^a\right)\derpar{}{g^{ab}} +x_b^\mu\partial_a\xi^b\frac{\partial}{\partial x_a^\mu} - \left(g^{ac}_d\partial_c \xi^b + g^{cb}_d\partial_c \xi^a\right)\derpar{}{g^{ab}_d} \in\mathfrak{X}(J^1\pi) \ ,
\end{equation*}
which is tangent to the final Lagrangian constraint submanifold $S_f$,
as a simple calculation shows.
The canonical lifts to the multimomentum bundles  are given 
by \eqref{liftJ1} and \eqref{liftJ1b}, respectively, and they are
\beann
Z_\xi&=&  -\xi^a\frac{\partial}{\partial\sigma^a} - \left(g^{ac}\partial_c \xi^b + g^{cb}\partial_c \xi^a\right)\derpar{}{g^{ab}} + \left(p^a_\mu\partial_b \xi^b-p^b_\mu\partial_b \xi^a\right)\derpar{}{p^a_\mu} \\ 
& & +\left(\pi^c_{ab}\partial_d\xi^d - \pi^d_{ab}\partial_d\xi^c + \pi^c_{ad}\partial_b\xi^d + \pi^c_{db}\partial_a\xi^d \right)\derpar{}{\pi^c_{ab}}
+p\partial_a\xi^a\derpar{}{p}\in\vf(\mathcal{M}\pi) \ , \\
Y_\xi&=& -\xi^a\frac{\partial}{\partial\sigma^a} - \left(g^{ac}\partial_c \xi^b + g^{cb}\partial_c \xi^a\right)\derpar{}{g^{ab}} + \left(p^a_\mu\partial_b \xi^b-p^b_\mu\partial_b \xi^a\right)\derpar{}{p^a_\mu} \\
& &\quad \quad \quad \quad 
 +\left(\pi^c_{ab}\partial_d\xi^d - \pi^d_{ab}\partial_d\xi^c + \pi^c_{ad}\partial_b\xi^d + \pi^c_{db}\partial_a\xi^d \right)\derpar{}{\pi^c_{ab}} \in\mathfrak{X}(J^1\pi^*)  \ .
\eeann
This vector field $Y_\xi$ is tangent to $P_\circ$, since the constraints defining $P_\circ$ are $\pi^c_{ab}=0$, and $\Lie(Y_\xi)\pi^c_{ab}=0$
on $P_\circ$.
Thus, following Definition \ref{liftvf0},
the canonical lift to the Hamiltonian multiphase space $P_\circ\subset J^1\pi^*$ is
given as
\begin{equation*}
Y^\circ_\xi = -\xi^a\frac{\partial}{\partial\sigma^a} - \left(g^{ac}\partial_c \xi^b + g^{cb}\partial_c \xi^a\right)\derpar{}{g^{ab}} + \left(p^a_\mu\partial_b \xi^b-p^b_\mu\partial_b \xi^a\right)\derpar{}{p^a_\mu} \in\mathfrak{X}(P_\circ)\ ,
\end{equation*}
since $({\jmath}_\circ)_*Y^\circ_\xi=Y_\xi\vert_{P_\circ}$.
It can be shown that $(\mathscr{FL}_\circ)_*X_\xi=Y_\xi^\circ$ is satisfied and that $Y_\xi^\circ$ is also tangent to the final Hamiltonian constraint submanifold $P_f$.

It is straightforward to show that $\Lie(X_\xi)\Omega_\mathscr{L}=0$;
then,
$$
0=\Lie(X_\xi)\Omega_\mathscr{L}=
\Lie(X_\xi)\big(\mathscr{FL}_\circ^{\ *}\,\Omega^\circ_\mathscr{H}\big)=
\mathscr{FL}_\circ^{\ *}\left(\Lie\big((\mathscr{FL}_\circ)_*X_\xi\big)\Omega^\circ_\mathscr{H}\right)=
\mathscr{FL}_\circ^{\ *}\big(\Lie(Y^\circ_\xi)\Omega^\circ_\mathscr{H}\big) \ ,
$$
and, as $\mathscr{FL}_\circ$ is a surjective submersion, we conclude that
$\Lie(Y^\circ_\xi)\Omega^\circ_\mathscr{H}=0$.
Hence, both $X_\xi$ and $Y^\circ_\xi$ are natural infinitesimal Noether symmetries.
The multimomentum maps associated with $X_\xi$ and $Y^\circ_\xi$ are given as,
\begin{align*}
J_\mathscr{L}(X_\xi)&=-i(X_\xi)\Theta_\mathscr{L} = T\sqrt{-g\,}\xi^c\eta_{\mu\nu}g^{ab}x^\nu_b \left(\frac{1}{2}x^\mu_a \text{d}^1\sigma_c-\epsilon_{ac}\text{d}x^\mu \right) \in\Omega^1(J^1\pi)  \ , \\
J^\circ_\mathscr{H}(Y^\circ_\xi) &=-i(Y^\circ_\xi)\Theta^\circ_\mathscr{H} = -\epsilon_{ab}\xi^b p^a_\mu\text{d}x^\mu - \frac{1}{2T\sqrt{-g\,}}
\eta_{\mu\nu}g_{ab}p^a_\mu p^b_\nu\xi^c\text{d}^1\sigma_c \in\Omega^1(P_\circ) \ .\
\end{align*} 

{\sl Poincar\'e transformations\/}:

Next, the infinitesimal Poincar\'e transformations of the background spacetime are given as $x^\mu\rightarrow x^\mu + \omega^\mu_{\ \nu}x^\nu + {\rm a}^\mu$ where both $\omega^\mu_{\ \nu}$ and ${\rm a}^\mu$ are constants and the matrices $\omega^\mu_{\ \nu}$ are antisymmetric, similarly to the Einstein--Cartan theory of the previous section. The vector field which generates these transformations on the configuration bundle $E$ is $\pi$-vertical and is given as,
\begin{equation*}
\zeta_E =-\left(\omega^\mu_{\ \nu}x^\nu + {\rm a}^\mu\right)\derpar{}{x^\mu}\ .
\end{equation*}
Its canonical lift \eqref{j1Y} to $J^1\pi$ is,
\begin{equation*}
X_\zeta = j^1\zeta_E = - \left(\omega^\mu_{\ \nu}x^\nu + {\rm a}^\mu\right)\derpar{}{x^\mu} 
 -x^\nu_a\,\omega^\mu_\nu\derpar{}{x^\mu_a}  \in\vf(J^1\pi)\ .
\end{equation*}
As expected, $X_\zeta$ is tangent to the final Lagrangian constraint submanifold $S_f$.
The canonical lifts to the multimomentum bundles, given respectively
by \eqref{liftJ1} and \eqref{liftJ1b}, are
\beann
Z_\zeta&=&-\left(\omega^\mu_{\ \nu}x^\nu + {\rm a}^\mu\right)\derpar{}{x^\mu}
+p^a_\nu\,\omega^\nu_\mu\derpar{}{p^a_\mu} \in\vf(\mathcal {M}\pi) \ , \\
Y_\zeta&=&-\left(\omega^\mu_{\ \nu}x^\nu + {\rm a}^\mu\right)\derpar{}{x^\mu}
+p^a_\nu\,\omega^\nu_\mu\derpar{}{p^a_\mu}\in \mathfrak{X}(J^1\pi^*) \ .
\eeann
As $\Lie(Y_\zeta)\pi^c_{ab}=0$ on $P_\circ$, it follows that $Y_\zeta$ is tangent to $P_\circ$ and, by Definition \ref{liftvf0},
the canonical lift to $P_\circ\subset J^1\pi^*$ is 
\begin{equation*}
Y^\circ_\zeta= - \left(\omega^\mu_{\ \nu}x^\nu + {\rm a}^\mu\right)\derpar{}{x^\mu}
+p^a_\nu\,\omega^\nu_\mu\derpar{}{p^a_\mu} 
\in\vf(P_\circ)\ .
\end{equation*}
As above, 
$(\mathscr{FL}_\circ)_*X_\zeta=Y_\zeta^\circ$,
and $Y_\zeta^\circ$ is also tangent to the final Hamiltonian constraint submanifold $P_f$ as expected.

Again $\Lie(X_\zeta)\Omega_\mathscr{L}=0$ and
$\Lie(Y^\circ_\zeta)\Omega^\circ_\mathscr{H}=0$;
$X_\zeta$ and $Y^\circ_\zeta$ are therefore natural infinitesimal Noether symmetries.
The corresponding multimomentum maps are:
\begin{align*}
J_\mathscr{L}(X_\zeta)&=-i(X_\zeta)\Theta_\mathscr{L}= - T\sqrt{-g\,}\eta_{\mu\rho}g^{ab} x^\rho_b \left(\omega^\mu_{\ \nu}x^\nu + {\rm a}^\mu\right)\text{d}^1\sigma_a \in\Omega^1(J^1\pi)\ , \\
J^\circ_\mathscr{H}(Y^\circ_\zeta)&= -i(Y^\circ_\zeta)\Theta^\circ_\mathscr{H}=p^a_\mu\left(\omega^\mu_{\ \nu}x^\nu + {\rm a}^\mu\right)\text{d}^1\sigma_a \in\Omega^1(P_\circ)\ .
\end{align*}

{\sl Weyl transformations\/}:

Finally, the Weyl gauge transformations are given as: 
$g^{ab}\rightarrow e^{-\varphi(\sigma)}g^{ab}$. 
These transformations are generated on the configuration manifold $E$ by the vector field $\varphi_E\in\vf(E)$ given as
\begin{equation*}
\varphi_E=e^{-\varphi}g^{ab}\derpar{}{g^{ab}}\ .
\end{equation*}
The canonical lift \eqref{j1Y} of $\varphi_E$ to the Lagrangian multiphase space $J^1\pi$ is,
\begin{equation*}
X_\varphi = j^1\varphi_E = e^{-\varphi}g^{ab}\derpar{}{g^{ab}} + e^{-\varphi}\left(g^{ab}_c-g^{ab}\partial_c \varphi\right)\derpar{}{g^{ab}_c}\in\vf(J^1\pi)\ , 
\end{equation*}
which is tangent to the final Lagrangian constraint submanifold $S_f$.
Furthermore, 
the canonical lifts \eqref{liftJ1} and \eqref{liftJ1b} to the multimomentum bundles are
\beann
Z_\varphi&=& - e^{-\varphi}g^{ab}\derpar{}{g^{ab}} + e^{-\varphi}\pi^c_{ab}\derpar{}{\pi^c_{ab}} - e^{-\varphi}g^{ab}\pi^c_{ab}\partial_c \varphi \derpar{}{p} 
\in\vf(\mathcal {M}\pi) \ , \\
Y_\varphi&=& - e^{-\varphi}g^{ab}\derpar{}{g^{ab}} + e^{-\varphi}\pi^c_{ab}\derpar{}{\pi^c_{ab}}
\in \mathfrak{X}(J^1\pi^*)\ .
\eeann
$Y_\varphi$ is tangent to $P_\circ$ so, by Definition \ref{liftvf0},
the canonical lift of $\varphi_E$ 
to the primary constraint submanifold $P_\circ\subset J^1\pi^*$ is 
\begin{equation*}
Y^\circ_\varphi = e^{-\varphi}g^{ab}\derpar{}{g^{ab}}\in\vf(P_\circ)\ .
\end{equation*}
Once again, it follows that
$(\mathscr{FL}_\circ)_*X_\varphi=Y_\varphi^\circ$, 
and $Y_\varphi^\circ$ is also tangent to the final Hamiltonian constraint submanifold $P_f$.

Once again, $\Lie(X_\varphi)\Omega_\mathscr{L}=0$ 
and
$\Lie(Y^\circ_\varphi)\Omega^\circ_\mathscr{H}=0$,
so it follows that $X_\varphi$ and $Y^\circ_\varphi$ are natural infinitesimal Noether symmetries. 
In addition, it is straightforward to show that
$i(X_\varphi)\Theta_\mathscr{L}=0$ and $i(Y^\circ_\varphi)\Theta^\circ_\mathscr{H}=0$, which means that their corresponding momentum maps are equal to zero.
This is what is expected in Weyl invariant field theories (see e.g. \cite{JP-2015}).

\subsection{Einstein--Cartan gravity in $3+1$ dimensions}
\label{Einstein-Cartan}

In this section, the premultisymplectic treatment of Einstein--Cartan gravity of $(3+1)$-dimensional spacetime $M$ is given. 
The vierbein $e^a_\mu$ and spin connection $\omega^{ab}_\mu$ are treated as independent variational fields.
The configuration manifold is the frame bundle augmented by the bundle of connections, given as 
$\pi:E\rightarrow M: (x^\mu, e^a_\mu, \omega^{ab}_\mu) \mapsto x^\mu$.
Note that the lowercase Latin indices are in the Lie algebra of the gauge group $SO(3,1)$ of local Lorentz transformations. 

Alternative treatments of Einstein--Cartan gravity in the multisymplectic setting exist (see e.g. \cite{Vey-2015}).
However, such works do not carry out a complete constraint analysis; in this work, we provide the full constraint analysis.
It is also worth noting that a different geometric treatment, sometimes referred to as the {\sl polysymplectic formalism}, exists;
for the treatment of vielbein gravity in this formalism, see e.g. \cite{Kanatchikov2}.

As above, the spacetime partial derivatives are denoted as $\partial_\mu = \partial/\partial x^\mu$.

\subsubsection{Lagrangian Formalism}

The {\sl Einstein--Hilbert Lagrangian} is written in terms of the tetrad $e^a_\mu$ and spin connection $\omega^a_{\mu c}$ as,
\begin{equation*}
\mathscr{L}= \epsilon^{\mu\nu\rho\sigma}\epsilon_{abcd}e^a_\mu e^b_\nu\left(\omega^{cd}_{\rho\sigma}+\omega^c_{\rho i}\omega^{id}_\sigma\right) \in C^\infty(J^1\pi)\ .
\end{equation*}
The multi-Hessian given by the second derivative of the Lagrangian with respect to the multivelocities $e^a_{\mu\nu}$ and $\omega^{ab}_{\mu\nu}$ vanishes and is thereby singular. 
The Lagrangian energy function $E_\mathscr{L}\in C^\infty(J^1\pi)$ is 
\begin{equation*}
E_\mathscr{L} = -\epsilon^{\mu\nu\rho\sigma}\epsilon_{abcd}e^a_\mu e^b_\nu \omega^c_{\rho i}\omega^{id}_\sigma\ ,
\end{equation*}
and the Cartan $4$-form and $5$-form are given as, 
\begin{align*}
\Theta_\mathscr{L} &=  \epsilon^{\mu\nu\rho\sigma}\epsilon_{abcd} e^a_\mu e^b_\nu \left(\text{d}\omega^{cd}_\sigma\wedge\text{d}^{3}x_\rho+\omega^c_{\rho i}\omega^{id}_\sigma\text{d}^4x\right)\ , \\
\Omega_\mathscr{L} &= -2\epsilon^{\mu\nu\rho\sigma}\epsilon_{abcd} \left(e^b_\nu\text{d}e^a_\mu\wedge\text{d}\omega^{cd}_\sigma\wedge\text{d}^{3}x_\rho + e^b_\nu\omega^c_{\rho i}\omega^{id}_\sigma\text{d}e^a_\mu\wedge\text{d}^4x - e^a_\mu e^i_\nu\omega^b_{\rho i} \text{d}\omega^{cd}_{\sigma}\wedge\text{d}^4x\right) \ .
\end{align*}

Similarly as above, to compute the field equations we use $4$-multivector fields expressed locally as,
\begin{equation}
{\bf X}_\mathscr{L} = \bigwedge^{3}_{\lambda=0}X_\lambda=\bigwedge^{3}_{\lambda=0}\left(\frac{\partial}{\partial x^\lambda}+B^a_{\lambda\mu}\frac{\partial}{\partial e^a_\mu}+C^c_{\lambda\rho d}\frac{\partial}{\partial \omega^c_{\rho d}}+D^a_{\lambda\sigma\mu}\frac{\partial}{\partial e^a_{\sigma\mu}}+H^c_{\lambda\sigma\rho d}\frac{\partial}{\partial\omega^c_{\sigma\rho d}}\right) \in \vf^4(J^1\pi)\ .
\end{equation}
Then $i({\bf X}_\mathscr{L})\Omega_\mathscr{L}=0$ gives:
\begin{align*}
0 &= \epsilon^{\mu\nu\rho\sigma}\epsilon_{abcd}e^b_\nu\left(C^{cd}_{\rho\sigma} + \omega^c_{\rho i}\omega^{id}_\sigma\right) \ ,
\\
0 &= \epsilon^{\mu\nu\rho\sigma}\epsilon_{abcd}e^b_\nu\left(B^a_{\rho\mu} + e^i_\mu\omega^a_{\rho i}\right) \ .
\end{align*}
Upon imposing the {\sc sopde} conditions, $B^a_{\rho\mu} =  e^a_{\rho\mu}$, $C^{cd}_{\rho\sigma} = \omega^{cd}_{\rho\sigma}$, 
the field equations become:
\begin{align*}
0 &= \epsilon^{\mu\nu\rho\sigma}\epsilon_{abcd}e^b_\nu\left(\omega^{cd}_{\rho\sigma} + \omega^c_{\rho i}\omega^{id}_\sigma\right) \ ,
\\
0 &= \epsilon^{\mu\nu\rho\sigma}\epsilon_{abcd}e^b_\nu\left( e^a_{\rho\mu} + e^i_\mu\omega^a_{\rho i}\right) \ .
\end{align*}
Both of these equations are {\sc sopde} constraints which specify the {\sc sopde} constraint submanifold. 
Imposing the stability (tangency) condition, by demanding that the Lie derivatives of the above constraints with respect to the vector fields $X_\lambda$ of ${\bf X}_\mathscr{L}$ vanish, yields relations for the ${\bf X}_\mathscr{L}$ multivector field coefficients $D^a_{\nu\sigma\mu}$ and $H^c_{\nu\sigma\rho d}$.
It therefore follows that no stability (tangency) constraints arise, and it follows that the {\sc sopde} constraint submanifold is the final constraint submanifold $S_f\hookrightarrow J^1\pi$.
This is shown component-wise as follows:
\beann
0&=&\left.\Lie(X_\lambda)\left[\epsilon^{\mu\nu\rho\sigma}\epsilon_{abcd}e^b_\nu\big(\omega^{cd}_{\rho\sigma} + \omega^c_{\rho i}\omega^{id}_\sigma\big)\right]\right\vert_{S_f} 
\\
&=&
\epsilon^{\mu\nu\rho\sigma}\epsilon_{abcd}\left[e^b_\nu\big(H^{cd}_{\lambda\rho\sigma} + \omega^c_{\lambda\rho i}\omega^{id}_\sigma + \omega^c_{\rho i}\omega^{id}_{\lambda\sigma}\big) + e^b_{\lambda\nu}\big(\omega^{cd}_{\rho\sigma} + \omega^c_{\rho i}\omega^{id}_\sigma\big)\right] \ ,
\\
0&=&\left.\Lie(X_\lambda)\left[\epsilon^{\mu\nu\rho\sigma}\epsilon_{abcd}e^b_\nu\big( e^a_{\rho\mu} + e^i_\mu\omega^a_{\rho i}\big)\right]\right\vert_{S_f}
\\
&=& \epsilon^{\mu\nu\rho\sigma}\epsilon_{abcd}\left[e^b_\nu\big( D^a_{\lambda\rho\mu} + e^i_{\lambda\mu}\omega^a_{\rho i} + e^i_\mu \omega^a_{\lambda\rho i}\big) + e^b_{\lambda\nu}\big( e^a_{\rho\mu} + e^i_\mu\omega^a_{\rho i}\big)\right]
\ .
\eeann 

Now, working on integral sections of $\textbf{X}_\mathscr{L}$,
the final field equations are given as,
\begin{align*}
0 &= \epsilon^{\mu\nu\rho\sigma}\epsilon_{abcd}e^b_\nu\left(\partial_\rho\omega^{cd}_\sigma + \omega^c_{\rho i}\omega^{id}_\sigma\right) \ ,
\\
0 &= \epsilon^{\mu\nu\rho\sigma}\epsilon_{abcd}e^b_\nu\left(\partial_\rho e^a_\mu + e^i_\mu\omega^a_{\rho i}\right) \ ,
\end{align*}
which are the well-known Euler--Lagrange equations for this theory.

\subsubsection{Hamiltonian Formalism}

Recall from the discussion in the Lagrangian setting that this theory of gravity is singular. 
The local coordinates on ${\cal M}\pi$ are $(x^\mu, e^a_\mu, \omega^{ab}_\mu,p^{\mu\nu}_a,\pi^{\rho\sigma}_{cd},p)$ and 
$J^1\pi^*$ has local coordinates $(x^\mu,e^a_\mu,\omega^{ab}_\mu,p^{\mu\nu}_a,\pi^{\rho\sigma}_{cd})$.
Furthermore, the primary constraint submanifold $P_\circ=\text{Im}\mathscr F\mathscr{L}\subset J^1\pi^*$ on which the Hamilton–-De Donder–-Weyl formalism 
takes place is given by the following equations:
\begin{equation}
\mathscr{FL}^*p^{\mu\nu}_a =\derpar{\mathscr{L}}{e^a_{\mu\nu}} = 0 \quad , \quad 
\mathscr{FL}^*\pi^{\rho\sigma}_{cd} = \derpar{\mathscr{L}}{\omega_{\rho\sigma}^{cd}} = \epsilon^{\mu\nu\rho\sigma}\epsilon_{abcd}e^a_\mu e^b_\nu \ .
\label{cons}
\end{equation}
The De Donder--Weyl Hamiltonian is, 
$$
\mathscr{H}_\circ = - \epsilon^{\mu\nu\rho\sigma}\epsilon_{abcd}e^a_\mu e^b_\nu \omega^c_{\rho i}\omega^{id}_\sigma \ ,
$$
and the Hamiltonian section is given as 
$h_\circ = (x^\mu, e^a_\mu, \omega^{ab}_\mu,p^{\mu\nu}_a,\pi^{\rho\sigma}_{cd},p=-\mathscr{H}_\circ)$.

The Hamilton--Cartan $4$-form and $5$-form are:
\begin{align*}
\Theta^\circ_\mathscr{H} &=  \epsilon^{\mu\nu\rho\sigma}\epsilon_{abcd} e^a_\mu e^b_\nu \left(\text{d}\omega^{cd}_\sigma\wedge\text{d}^{3}x_\rho+\omega^c_{\rho i}\omega^{id}_\sigma\text{d}^4x\right)\ , \\
\Omega^\circ_\mathscr{H} &= -2\epsilon^{\mu\nu\rho\sigma}\epsilon_{abcd} \left(e^b_\nu\text{d}e^a_\mu\wedge\text{d}\omega^{cd}_\sigma\wedge\text{d}^{3}x_\rho + e^b_\nu\omega^c_{\rho i}\omega^{id}_\sigma\text{d}e^a_\mu\wedge\text{d}^4x - e^a_\mu e^i_\nu\omega^b_{\rho i} \text{d}\omega^{cd}_{\sigma}\wedge\text{d}^4x\right) \ .
\end{align*}
Now using a $4$-multivector field  with components,
\begin{equation}
{\bf X}^\circ_\mathscr{H} = \bigwedge^{3}_{\nu=0} X_\nu=\bigwedge^{3}_{\mu=0} \left(\frac{\partial}{\partial x^\nu}+B^a_{\nu\mu}\frac{\partial}{\partial e^a_\mu}+C^c_{\nu\rho d}\frac{\partial}{\partial \omega^c_{\rho d}}\right) \in\vf^4(P_\circ) \ ,
\end{equation}
the equation $i({\bf X}^\circ_\mathscr{H})\Omega^\circ_\mathscr{H}=0$ gives,
\begin{align*}
0 &= \epsilon^{\mu\nu\rho\sigma}\epsilon_{abcd}e^b_\nu\left(C^{cd}_{\rho\sigma} + \omega^c_{\rho i}\omega^{id}_\sigma\right) \ ,
\\
0 &= \epsilon^{\mu\nu\rho\sigma}\epsilon_{abcd}e^b_\nu\left(B^a_{\rho\mu} + e^i_\mu\omega^a_{\rho i}\right) \ ,
\end{align*}
and evidently no compatibility constraints arise.
Note that the {\sc sopde} constraints from the Lagrangian setting are not $\mathscr{FL}$-projectable (see \cite{AGR-2022}). 
Furthermore, as in the Lagrangian setting, there are no compatibility constraints. 
The final field equations are given on integral sections of ${\bf X}^\circ_\mathscr{H}$, and are written as
\begin{align*}
0 &= \epsilon^{\mu\nu\rho\sigma}\epsilon_{abcd}e^b_\nu\left(\partial_\rho\omega^{cd}_\sigma + \omega^c_{\rho i}\omega^{id}_\sigma\right) \ ,
\\
0 &= \epsilon^{\mu\nu\rho\sigma}\epsilon_{abcd}e^b_\nu\left(\partial_\rho e^a_\mu + e^i_\mu\omega^a_{\rho i}\right) \ ,
\end{align*}
identically to the Lagrangian formulation.

\subsubsection{Symmetries}

The two kinds of symmetries exhibited by this theory are {\sl spacetime diffeomorphisms} and local {\sl Poincar\'e transformations}. 

{\sl Spacetime diffeomorphisms\/}:

Take spacetime diffeomorphisms to be generated infinitesimally by the vector field,
\begin{equation}
\xi = -\xi^\mu(x)\derpar{}{x^\mu}\in\vf(M) \ ,
\end{equation}
by performing (point-wise) the following coordinate transformation:
$x^\mu \rightarrow x^\mu + \xi^\mu(x)$.
It follows that the fields of the theory transform as,
\beann
\delta e^a_\mu(x)=e'^a_\mu(x)-e^a_\mu(x) =-\xi^\nu\partial_\nu e^a_\mu-e^a_\nu\partial_\mu \xi^\nu \ , \\
\delta \omega^a_{\mu b}(x)=\omega'^a_{\mu b}(x)-\omega^a_{\mu b}(x)=-\xi^\nu\partial_\nu \omega^a_{\mu b}-\omega^a_{\nu b}\partial_\mu \xi^\nu \ .
\eeann
The canonical lift to the configuration bundle  $E$ is given by,
\begin{equation*}
\xi_E= -\xi^\mu \derpar{}{x^\mu} + e^a_\nu \partial_\mu \xi^\nu \frac{\partial }{\partial e^a_\mu} + \omega^a_{\nu b}\partial_\mu \xi^\nu\frac{\partial}{\partial \omega^a_{\mu b}} \in \mathfrak{X}(E) \ .
\end{equation*}
As usual, the canonical lift of $\xi_E$ to the Lagrangian phase space $J^1\pi$ is the $1$-jet prolongation of $\xi_E$ given by \eqref{j1Y} as
\begin{align}
\begin{split}
X_\xi = j^1\xi_E = &-\xi^\mu \derpar{}{x^\mu}+ e^a_\nu \partial_\mu \xi^\nu\frac{\partial }{\partial e^a_\mu}+ \omega^a_{\nu b}\partial_\mu \xi^\nu\frac{\partial}{\partial \omega^a_{\mu b}} \\
&+\left(e^a_{\sigma\rho}\partial_\mu \xi^\rho-e^a_{\rho\mu}\partial_\sigma \xi^\rho\right)\frac{\partial}{\partial e^a_{\sigma\mu}}
+\left(\omega^a_{\sigma\rho b}\partial_\mu \xi^\rho-\omega^a_{\rho\mu b}\partial_\sigma \xi^\rho\right)\frac{\partial}{\partial \omega^a_{\sigma\mu b}} \in\vf(J^1\pi)\ .
\end{split}
\end{align}
and, by \eqref{liftJ1} and \eqref{liftJ1b}, 
the canonical lifts to the multimomentum bundles  are
\beann
Z_\xi&=&-\xi^\mu \derpar{}{x^\mu}+ e^a_\nu \partial_\mu \xi^\nu\frac{\partial }{\partial e^a_\mu}+ \omega^a_{\nu b}\partial_\mu \xi^\nu\frac{\partial}{\partial \omega^a_{\mu b}} \\
& &- \left(\partial_\lambda\xi^\mu p^{\lambda\nu}_a - \partial_\lambda\xi^\lambda p^{\mu\nu}_a - \partial_\lambda\xi^\nu p^{\mu\lambda}_a\right)\derpar{}{p^{\mu\nu}_a} \\
& &- \left(\partial_\lambda\xi^\mu \pi^{\lambda\nu}_{ab} - \partial_\lambda\xi^\lambda \pi^{\mu\nu}_{ab} - \partial_\lambda\xi^\nu \pi^{\mu\lambda}_{ab}\right)\derpar{}{\pi^{\mu\nu}_{ab}}
+ p\,\partial_\mu\xi^\mu\derpar{}{p}  \in\vf(\mathcal{M}\pi) \ , \\
Y_\xi&=& -\xi^\mu \derpar{}{x^\mu}+ e^a_\nu \partial_\mu \xi^\nu\frac{\partial }{\partial e^a_\mu}+ \omega^a_{\nu b}\partial_\mu \xi^\nu\frac{\partial}{\partial \omega^a_{\mu b}} \\ 
& &- \left(\partial_\lambda\xi^\mu p^{\lambda\nu}_a - \partial_\lambda\xi^\lambda p^{\mu\nu}_a - \partial_\lambda\xi^\nu p^{\mu\lambda}_a\right)\derpar{}{p^{\mu\nu}_a} \\
& &- \left(\partial_\lambda\xi^\mu \pi^{\lambda\nu}_{ab} - \partial_\lambda\xi^\lambda \pi^{\mu\nu}_{ab} - \partial_\lambda\xi^\nu \pi^{\mu\lambda}_{ab}\right)\derpar{}{\pi^{\mu\nu}_{ab}}
\in\mathfrak{X}(J^1\pi^*) \ .
\eeann
In this theory, it can be shown that
$(\mathscr{FL}_\circ)_*X_\xi=Y_\xi^\circ\in\vf(P_\circ)$, where
\begin{align}
\begin{split}
Y^\circ_\xi = -\xi^\mu \derpar{}{x^\mu}+ e^a_\nu \partial_\mu \xi^\nu\frac{\partial }{\partial e^a_\mu}+ \omega^a_{\nu b}\partial_\mu \xi^\nu\frac{\partial}{\partial \omega^a_{\mu b}} \in\vf(P_\circ)\ .
\end{split}
\end{align}
This vector field is not the canonical lift of $\xi_E$ to $P_\circ$ according to Definition \ref{liftvf0} 
because the Lie derivatives by $Y_\xi$ of the primary constraint functions which arise from \eqref{cons} do not vanish on $P_\circ$, and hence,
$Y_\xi$ is not tangent to $P_\circ$.

Nevertheless, observe that, as $\Lie(X_\xi)\Omega_\mathscr{L}=0$, it follows that $\Lie(Y^\circ_\xi)\Omega^\circ_\mathscr{H}=0$ as in the previous example.
$X_\xi$ is a natural Lagrangian infinitesimal Noether symmetry but $Y^\circ_\xi$ is a Hamiltonian infinitesimal Noether symmetry which is not natural according to our Definition
\ref{Cartansym0},
although it comes from a symmetry in $E$.
The multimomentum maps associated with $X_\xi$ and $Y^\circ_\xi$ are:
\begin{align*}
J_\mathscr{L}(X_\xi)&=-i(X_\xi)\Theta_\mathscr{L} = \epsilon^{\mu\nu\rho\sigma}\epsilon_{abcd} e^a_\mu e^b_\nu \xi^\lambda\left(\omega^c_{\rho i}\omega^{id}_\sigma\text{d}^3x_\lambda-\text{d}\omega^{cd}_\sigma\wedge\text{d}^{2}x_{\rho\lambda}\right)\ ,
\\
J^\circ_\mathscr{H}(Y^\circ_\xi) &=-i(Y^\circ_\xi)\Theta^\circ_\mathscr{H} = \epsilon^{\mu\nu\rho\sigma}\epsilon_{abcd} e^a_\mu e^b_\nu \xi^\lambda\left(\omega^c_{\rho i}\omega^{id}_\sigma\text{d}^3x_\lambda-\text{d}\omega^{cd}_\sigma\wedge\text{d}^{2}x_{\rho\lambda}\right)\ .
\end{align*} 

{\sl Local Lorentz transformations\/}:

The $SO(3,1)$ gauge group of local Lorentz transformations acts point-wise on the tangent space coordinates $y^a$ infinitesimally as
\begin{equation*}
y^a\to y^a - \Lambda^a_{\ b}y^b\ ,
\end{equation*}
where the infinitesimal Lorentz matrices $\Lambda^a_{\ b}$ are skewsymmetric.

The local Lorentz transformations of the fields are given as,
\begin{equation}
\delta e^a_\mu = \Lambda^a_{\ b}e^b_\mu \quad , \quad
\delta \omega^a_{\mu b} = \Lambda^a_{\ c} \left(\Lambda^{-1}\right)^d_{\ b} \omega^c_{\mu d} + \Lambda^a_{\ c} \partial_\mu \left(\Lambda^{-1}\right)^c_{\ b} \ . 
\end{equation}
The vector field on the configuration bundle $E$ which generates these transformations is given as:
\begin{equation*}
\zeta_E=\Lambda^a_{\ b}e^b_\mu \derpar{}{e^a_\mu} + \left[\Lambda^a_{\ c} \left(\Lambda^{-1}\right)^d_{\ b} \omega^c_{\mu d} + \Lambda^a_{\ c} \partial_\mu \left(\Lambda^{-1}\right)^c_{\ b}\right]\derpar{}{\omega^a_{\mu b}} \in \vf(E) \ .
\end{equation*}
The canonical lift of $\zeta_E$ to the multivelocity phase space $J^1\pi$ is 
\begin{align*}
\begin{split}
X_\zeta &=j^1\zeta_E = \Lambda^a_{\ b}e^b_\mu \derpar{}{e^a_\mu} + \left[\Lambda^a_{\ c} \left(\Lambda^{-1}\right)^d_{\ b} \omega^c_{\mu d} + \Lambda^a_{\ c} \partial_\mu \left(\Lambda^{-1}\right)^c_{\ b}\right]\derpar{}{\omega^a_{\mu b}} + \left(e^b_\mu\partial_\rho \Lambda^a_{\ b} + e^b_{\rho\mu}\Lambda^a_{\ b}\right)\derpar{}{e^a_{\rho\mu}} 
\\
& + \left\{\omega^c_{\mu d}\partial_\rho\left[\Lambda^a_{\ c}\left(\Lambda^{-1}\right)^d_{\ b}\right] + \partial_\rho \left[\Lambda^a_{\ c}\partial_\mu \left(\Lambda^{-1}\right)^c_{\ b}\right] + \omega^c_{\rho\mu d}\Lambda^a_{\ c}\left(\Lambda^{-1}\right)^d_{\ b}\right\}\derpar{}{\omega^a_{\rho\mu b}} \in\vf(J^1\pi) \ .
\end{split}
\end{align*}
The local expressions for the canonical lifts \eqref{liftJ1} and \eqref{liftJ1b} 
 to the multimomentum bundles give
\beann
Z_\zeta&=&\Lambda^a_{\ b}e^b_\mu \derpar{}{e^a_\mu} + \left[\Lambda^a_{\ c} \left(\Lambda^{-1}\right)^d_{\ b} \omega^c_{\mu d} + \Lambda^a_{\ c} \partial_\mu \left(\Lambda^{-1}\right)^c_{\ b}\right]\derpar{}{\omega^a_{\mu b}}
\\ & & 
- \Lambda^b_{\ a} p^{\rho\mu}_b \derpar{}{p^{\rho\mu}_a} - \Lambda^c_{\ a}\left(\Lambda^{-1}\right)_b^{\ d} \pi^{\rho\mu}_{cd} \derpar{}{\pi^{\rho\mu}_{ab}} \\ & & 
-\left[ e^b_\mu p^{\rho\mu}_a \partial_\rho \Lambda^a_{\ b}
+ \omega^c_{\mu d} \pi^{\rho\mu}_{ab}\partial_\rho \Big(\Lambda^a_{\ c}(\Lambda^{-1})^{db} \Big) 
+ \pi^{\rho\mu}_{ab}\partial_\rho \Lambda^a_{\ c}\partial_\mu (\Lambda^{-1})^{cb} \right] \derpar{}{p} \in\vf(\mathcal{M}\pi)  \ , \\
Y_\zeta&=& \Lambda^a_{\ b}e^b_\mu \derpar{}{e^a_\mu} + \left[\Lambda^a_{\ c} \left(\Lambda^{-1}\right)^d_{\ b} \omega^c_{\mu d} + \Lambda^a_{\ c} \partial_\mu \left(\Lambda^{-1}\right)^c_{\ b}\right]\derpar{}{\omega^a_{\mu b}} \\
& & 
- \Lambda^b_{\ a} p^{\rho\mu}_b \derpar{}{p^{\rho\mu}_a} - \Lambda^c_{\ a}\left(\Lambda^{-1}\right)_b^{\ d} \pi^{\rho\mu}_{cd} \derpar{}{\pi^{\rho\mu}_{ab}}\in\mathfrak{X}(J^1\pi^*) \ .
\eeann
As before, 
$(\mathscr{FL}_\circ)_*X_\zeta=Y_\zeta^\circ$ , where now
\begin{align}
\begin{split}
Y^\circ_\zeta = \Lambda^a_{\ b}e^b_\mu \derpar{}{e^a_\mu} + \left[\Lambda^a_{\ c} \left(\Lambda^{-1}\right)^d_{\ b} \omega^c_{\mu d} + \Lambda^a_{\ c} \partial_\mu \left(\Lambda^{-1}\right)^c_{\ b}\right]\derpar{}{\omega^a_{\mu b}} \in\vf(P_\circ)\ .
\end{split}
\end{align}
Both $X_\zeta$ and $Y^\circ_\zeta$ are infinitesimal Noether symmetries as 
$\Lie(X_\zeta)\Omega_\mathscr{L}=0$ and
$\Lie(Y^\circ_\zeta)\Omega^\circ_\mathscr{H}=0$; again,
$X_\zeta$ is natural while $Y^\circ_\zeta$ is not natural 
as it does not satisfy the tangency condition of 
Definition \ref{Cartansym0}. 
The multimomentum maps associated with $X_\zeta$ and $Y^\circ_\zeta$ are:
\begin{align*}
\begin{split}
J_\mathscr{L}(X_\zeta) &= -i(X_\zeta)\Theta_\mathscr{L} = - \epsilon^{\mu\nu\rho\sigma}\epsilon_{abcd} e^a_\mu e^b_\nu \left[\Lambda^c_{\ i}\left(\Lambda^{-1}\right)^{jd} \omega^i_{\sigma j} + \Lambda^c_{\ j}\partial_\sigma \left(\Lambda^{-1}\right)^{jd}\right]
\text{d}^{3}x_\rho \ ,
\end{split}
\\
\begin{split}
J^\circ_\mathscr{H}(Y^\circ_\zeta) &= -i(Y^\circ_\zeta)\Theta^\circ_\mathscr{H} = - \epsilon^{\mu\nu\rho\sigma}\epsilon_{abcd} e^a_\mu e^b_\nu \left[\Lambda^c_{\ i}\left(\Lambda^{-1}\right)^{jd} \omega^i_{\sigma j} + \Lambda^c_{\ j}\partial_\sigma \left(\Lambda^{-1}\right)^{jd}\right]
\text{d}^{3}x_\rho \ .
\end{split}
\end{align*}

\section{Conclusions and outlook}

In this work we have introduced 
the intrinsic definition of the canonical lift of diffeomorphisms, and their infinitesimal generating vector fields, of the configuration manifold of a first--order classical field theory 
to the extended and restricted multimomentum bundles, 
${\cal M}\pi$ and $J^1\pi^*$.
We have also presented the appropriate restricted definition for such canonical lifts relevant to singular (or, more specifically, almost-regular) field theories which are formulated on embedded constraint submanifolds of the multimomentum bundles 
${\cal M}\pi$ and $J^1\pi^*$.
In general, the definitions we have given for the aforementioned canonical lifts depend only on the canonical multisymplectic structure of the extended multimomentum bundle ${\cal M}\pi$. 
This provides us with a powerful geometric tool to investigate field theories in the De Donder--Weyl Hamiltonian setting without having to make any reference to a Lagrangian formulation.
It should be noted that, in this work, we have outlined for the reader with the (pre)multisymplectic Lagrangian formulation of classical field theories with the purpose of pointing out the equivalence of the well-known canonical lifts in the Lagrangian setting \cite{book:Saunders89} to the canonical lifts in the De Donder--Weyl Hamiltonian setting which we introduce in this work. 

It is interesting to observe that the last field theory studied in this paper in Section \ref{Einstein-Cartan}, Einstein--Cartan gravity, 
is an example of a case in which we have Hamiltonian infinitesimal Noether symmetries which are not natural in the way that is presented in Definition \ref{Cartansym0}.
This is because the canonical lifts of the corresponding vector fields are not tangent to the primary constraint submanifold $P_\circ$, although they really come from infinitesimal symmetries originating from the configuration space $E$.
Interestingly, the symmetries of Enstein--Cartan gravity in the Hamiltonian setting are the Legendre map projection of Noether symmetries which are natural in the Lagrangian setting.
This scenario may arise in other field theories as well and suggests the possibility of establishing a more general definition of canonical lifts for almost-regular Hamiltonian field theories
as a part of a deeper study of symmetries for singular field theories in the De Donder--Weyl Hamiltonian formalism.

Being able to work directly on the multimomentum bundles ${\cal M}\pi$ and $J^1\pi^*$  without making reference to a Lagrangian formulation provides us with the ability to not only investigate field theories which may not have a Lagrangian description, but also search for new physical field theories completely geometrically in a new way.

\appendix

\section{Multivector fields on manifolds and fiber bundles}
\label{append}

(See \cite{Ca96a,CIL99,art:Echeverria_Munoz_Roman98}).
Let $\mathcal{M}$ be an $N$-dimensional differentiable manifold.
The {\sl \textbf{$m$-multivector fields}} in $\mathcal{M}$ ($m\leq N$)
are the sections of the $m$-multitangent bundle
$\displaystyle\bigwedge^m\Tan\mathcal{M}:=\overbrace{\Tan\mathcal{M}\wedge\ldots\wedge\Tan\mathcal{M}}^m$; that is,
the skew-symmetric contravariant tensor fields of order $m$ in $\mathcal{M}$;
the set of which is denoted $\vf^m (\mathcal{M})$.
Then, if $\mathbf{X}\in\mathfrak{X}^m(\mathcal{M})$,
for every point $\bar y\in \mathcal{M}$,
there is an open neighbourhood $U\subset \mathcal{M}$ and
$X_1,\ldots ,X_r\in\mathfrak{X} (U)$ such that,
for $m \leqslant r\leqslant{\rm dim}\,\mathcal{M}$,
$$
\mathbf{X}\vert_{U}=\sum_{1\leq i_1<\ldots <i_m\leq r} f^{i_1\ldots i_m}X_{i_1}\wedge\ldots\wedge X_{i_m} \, ,
$$
with $f^{i_1\ldots i_m} \in C^\infty (U)$.
In particular, $\mathbf{X}\in\vf^m(\mathcal{M})$ is said to be a
{\sl \textbf{locally decomposable multivector field}} if
there exist $X_1,\ldots ,X_m\in\vf (U)$ such that $\mathbf{X}\vert_U=X_1\wedge\ldots\wedge X_m$.
The locally decomposable $m$-multivector fields are locally associated with $m$-dimensional
distributions $D\subset\Tan \mathcal{M}$, and this splits
$\vf^m (\mathcal{M})$ into {\sl equivalence classes} $\{ {\bf X}\}\subset\vf^m(\mathcal{M})$ 
which are made of the locally decomposable multivector fields associated with the same distribution.
If ${\bf X},{\bf X}'\in\{ {\bf X}\}$ then, for $U\subset \mathcal{M}$,
there exists a non-vanishing function $f\in C^\infty(U)$ such that 
${\bf X}'=f{\bf X}$ on $U$.

If $\Omega\in\df^p(\mathcal{M})$ and $\mathbf{X}\in\mathfrak{X}^m(\mathcal{M})$,
the {\sl \textbf{contraction}} between ${\bf X}$ and $\Omega$ is
the natural contraction between tensor fields; in particular,
it gives zero when $p<m$ and, if $p\geq m$,
$$
 \inn({\bf X})\Omega\mid_{U}:= \sum_{1\leq i_1<\ldots <i_m\leq
 r}f^{i_1\ldots i_m} \inn(X_1\wedge\ldots\wedge X_m)\Omega 
=
 \sum_{1\leq i_1<\ldots <i_m\leq r}f^{i_1\ldots i_m} \inn
 (X_m)\ldots\inn (X_1)\Omega \ .
$$
The {\sl \textbf{Lie derivative}} of $\Omega$ with respect to ${\bf X}$ is the graded bracket (of degree $m-1$)
 $$
\Lie({\bf X})\Omega:=[\d , \inn ({\bf X})]\Omega=
(\d\inn ({\bf X})-(-1)^m\inn ({\bf X})\d)\Omega \ .
 $$

Now, let $\varrho\colon\mathcal{M}\to M$ a fiber bundle.
A multivector field $\mathbf{X}\in\mathfrak{X}^m(\mathcal{M})$ 
is \emph{$\varrho$-transverse} if,
for every $\beta\in\Omega^m(M)$ such that
$\beta_{\varrho({\rm p})}\not= 0$,
at every point ${\rm p}\in \mathcal{M}$,
we have that
$(\inn(\mathbf{X})(\varrho^*\beta))_{{\rm p}}\not= 0$.
We are interested in the {\sl \textbf{integrable multivector fields}},
which are those locally decomposable multivector fields whose associated distribution is integrable.
Then, if $\mathbf{X}\in\mathfrak{X}^m(\mathcal{M})$ is
integrable and $\varrho$-transverse, 
its integral manifolds are local sections of the projection
$\varrho\colon \mathcal{M}\to M$.

Furthermore, the {\sl \textbf{canonical prolongation}} of a section 
$\psi\colon U\subset M\to \mathcal{M}$ to 
$\Lambda^m\Tan{\cal M}$ is
the section $\psi^{(m)}\colon U\subset M\to\Lambda^m\Tan\mathcal{M}$ 
defined as $\psi^{(m)}:=\Lambda^m\Tan\psi\circ{\bf Y}_\omega$ where 
$\Lambda^m\Tan\psi\colon\Lambda^m\Tan M\to\Lambda^m\Tan\mathcal{M}$ is the natural extension of $\psi$ 
to the corresponding multitangent bundles
and ${\bf Y}_\omega\in\vf^m(M)$ is the unique $m$-multivector field on $M$
such that $\inn({\bf Y}_\omega)\omega=1$. Then,
$\psi$ is an integral section of ${\bf X}\in\vf^m({\cal M})$ if, and only if, $\psi^{(m)}={\bf X}\circ\psi$.

\section*{Acknowledgments}

We acknowledge the financial support of the 
{\sl Ministerio de Ciencia, Innovaci\'on y Universidades} (Spain), projects PID2021-125515NB-C21, and RED2022-134301-T of AEI,
and Ministry of Research and Universities of
the Catalan Government, project 2021 SGR 00603 \textsl{Geometry of Manifolds and Applications, GEOMVAP}.
We thank Profs. Joaquim Gomis and Manuel de Le\'on for enlightening discussions.
Our thanks also to the referees for their accurate comments and corrections that have allowed us to substantially improve the final version of the work.


\begin{thebibliography}{99}

{\small

\bibitem{AGR-2022}
D. Adame-Carrillo, J. Gaset, and N. Rom\'an-Roy,
``The second-order problem for k-presymplectic Lagrangian field theories: application to the Einstein–Palatini model'',
{\sl RACSAM} {\bf 116}, 20 (2022). (\url{https://doi.org/10.1007/s13398-021-01136-x}).

\bibitem{art:Aldaya_Azcarraga78_2}
V. Aldaya and J.A. de Azcarraga,
``Variational Principles on $rth$ order jets of fiber bundles in Field Theory'', 
\textsl{J. Math. Phys.} {\bf 19}(9) (1978) 1869--1875.
(\url{https://doi.org/10.1063/1.523904}).

\bibitem{AA-78}
V. Aldaya and J.A. de Azc\'arraga,
``Vector bundles, $rth$-order Noether invariants and canonical symmetries in Lagrangian field theory'',
{\sl J. Math. Phys.} {\bf 19}(9) (1978) 1876--1880.
(\url{https://doi.org/10.1063/1.523905}).

\bibitem{AA-80}
V. Aldaya and J.A. de Azc\'arraga,
``Geometric formulation of classical mechanics and field theory'',
{\sl Riv. Nuovo Cim.} {\bf 3} (1980) 1--66.
(\url{https://doi.org/10.1007/BF02906204}).

\bibitem{Gomis-String}
C. Batlle, J. Gomis, and J.M. Pons
``Hamiltonian and Lagrangian Constraints of the Bosonic String'',
{\sl Phys.Rev.D} {\bf 34} (1986) 2430--2432.
(\url{https://doi.org/10.1103/PhysRevD.34.2430}).

\bibitem{BBS}
K. Becker, M. Becker, and J.H. Schwarz,
{\it String Theory and M-Theory. A Modern Introduction},
Cambridge Univ. Press, Cambridge, 2006.
(\url{https://doi.org/10.1017/CBO9780511816086}).

\bibitem{Ca96a}
F. Cantrijn, L.A. Ibort, and M. de Le\'on,
``Hamiltonian structures on multisymplectic manifolds'',
{\sl Rend. Sem. Mat. Univ. Pol. Torino} \textbf{54}(3) (1996) 225--236.
(\url{https://www.researchgate.net/publication/233917958}).

\bibitem{CIL99}
F. Cantrijn, L.A. Ibort, and M. de Le\'on,
``On the geometry of multisymplectic manifolds'',
{\sl J. Austral. Math. Soc. (Series A)} \textbf{66}(3) (1999) 303--330.
(\url{https://www.researchgate.net/publication/231926129}).

\bibitem{CCI-91}
J.F. Cari\~nena, M. Crampin, and L.A. Ibort,
``On the multisymplectic formalism for first order field theories'',
{\sl Diff. Geom. Appl.} {\bf 1}(4) (1991) 345--374.
(\url{https://doi.org/10.1016/0926-2245(91)90013-Y}).

\bibitem{dD-1930}
T. de Donder, 
\emph{Th\'eorie invariantive du calcul des variations}.
Gauthier-Villars, Paris, 1935.

\bibitem{LMM-96}
{\rm M. de Le\'on, J. Mar\'\i n-Solano, and J.C. Marrero},
``A Geometrical approach to Classical Field Theories:  A constraint algorithm for singular theories'',
{\sl Proc.  New Develops. Diff. Geom.},
L. Tamassi, J. Szenthe eds., Kluwer Acad. Press, Amsterdam (1996) 291--312.
(\url{https://doi.org/10.1007/978-94-009-0149-0_
22}).

\bibitem{LMMMR-2005}
{\rm M. de Le\'on, J. Mar\'\i n-Solano, J.C. Marrero,
 M.C. Mu\~noz-Lecanda, and N. Rom\'an-Roy},
``Premultisymplectic constraint algorithm for field theories'',
{\sl Int. J. Geom. Meth. Mod. Phys.} {\bf 2} (2005) 839--871. 
(\url{http://doi.org/10.1142/S0219887805000880}).

\bibitem{LMD-2002}
M. de Leon, J.C. Marrero, D. Martin de Diego,
``A new geometric setting for classical field theories'',
{\sl Banach Center Pub. Inst. Math.}, Polish Academy of Sciences {\bf 59} (2003), 189-209.
(\url{http://doi.org/10.4064/bc59-0-10}).

\bibitem{art:deLeon_etal2004}
M. de Le\'on, D. Mart\'in de Diego, and A. Santamar\'ia-Merino,
``Symmetries in classical field theory'',
{\sl Int. J. Geom. Meths. Mod. Phys.} {\bf 1}(5) (2004) 651--710. 
(\url{http://doi.org/10.1142/S0219887804000290}).

\bibitem{De-77}
P. Dedecker,
``On the generalization of symplectic geometry to multiple integrals in the calculus of variations'',
in {\sl Differential Geometrical Methods in Mathematical Physics} (Proc. Sympos., Univ. Bonn, Bonn, 1975)
{\sl Lecture Notes in Math.} {\bf 570}, Springer, Berlin, 1977, 395--456.
(\url{https://doi.org/10.1007/BFb0087794}).

\bibitem{EMR-96}
A. Echeverr\'ia-Enr\'iquez, M.C. Mu\~noz-Lecanda, and N. Rom\'an-Roy,
``Geometry of Lagrangian first-order classical field theories'',
{\sl Forts. Phys.} {\bf 44} (1996) 235--280.
(\url{https://doi.org/10.1002/prop.2190440304}).

\bibitem{art:Echeverria_Munoz_Roman98}
A. Echeverr\'ia-Enr\'iquez, M.C. Mu\~noz-Lecanda, and N. Rom\'an-Roy,
``Multivector fields and connections: Setting Lagrangian equations in field theories'',
{\sl J. Math. Phys.} \textbf{39}(9) (1998) 4578--4603.
(\url{https://doi.org/10.1063/1.532525}).

\bibitem{EMR-99b}
A. Echeverr\'ia-Enr\'iquez, M.C. Mu\~noz-Lecanda, and N. Rom\'an-Roy, 
 ``Multivector field formulation of Hamiltonian
 field theories: Equations and symmetries'',
 {\sl J. Phys. A: Math. Gen.} {\bf 32} (1999) 8461--8484.
 (\url{https://doi.org/10.1088/0305-4470/32/48/309}).

\bibitem{EMR-00b}
A. Echeverr\'ia-Enr\'iquez, M.C. Mu\~noz-Lecanda, N. Rom\'an-Roy,
``Geometry of multisymplectic Hamiltonian first-order field theories'',
{\sl J. Math. Phys.} {\bf 41}(11) (2000) \textsl{J. Math. Phys.} {\bf 41}(11) (2000) 7402-7444.
 (\url{https://doi.org/10.1063/1.1308075}).

 \bibitem{FS-2012}
M. Forger and B.L. Soares,
``Local symmetries in gauge theories in a finite-dimensional setting'',
{\sl J. Geom. Phys.} {\bf 62}(9) (2012) 1925--1938. 
(\url{https://doi.org/10.1016/j.geomphys.2012.05.003}).

\bibitem{Gc-73}
P.L. Garc\'ia,
``The Poincar\'e-Cartan invariant in the calculus of variations'',
{\sl Symp. Math.} {\bf 14} (1973) 219--246.

\bibitem{art:GPR-2016}
J. Gaset, P.D. Prieto-Mart\'inez, and N. Rom\'an-Roy,
``Variational principles and symmetries on fibered multisymplectic manifolds'',
{\sl Comm. Math.} {\bf 24}(2) 137-152.
(\url{https://doi.org/10.1515/cm-2016-0010}).

\bibitem{GR1}
J. Gaset, N. Rom{\'a}n-Roy.
``Multisymplectic unified formalism for Einstein-Hilbert Gravity'',
 \emph{J. Math. Phys.} {\bf 59}(3) (2018) 032502.
(\url{https://doi.org/10.1063/1.4998526}).

\bibitem{GR2}
J. Gaset, N. Rom{\'a}n-Roy,
``New multisymplectic approach to the Metric-Affine (Einstein--Palatini) action for gravity''.
{\it J. Geom. Mech.} {\bf 11}(3) (2019) 361-396. 
(\url{https://doi.org/10.3934/jgm.2019019}).

\bibitem{GR-3}
J. Gaset and N. Rom\'an-Roy,
``Symmetries and gauge symmetries in multisymplectic first and second-order Lagrangian field theories: electromagnetic and gravitational fields'',
{\sl Rev. Acad.  Ciencias de Canarias (RACC)} {\bf 32}(2) 2020 (2021) 61--84.

\bibitem{GMS-97}
G. Giachetta, L. Mangiarotti, and G. Sardanashvily,
{\it New Lagrangian and Hamiltonian methods in field theory}, 
World Scientific Publishing Co., Inc., River Edge, NJ, 1997.
(\url{https://doi.org/10.1142/2199}).

\bibitem{GS-73}
H. Goldschmidt and S. Sternberg,
``The Hamilton--Cartan formalism in the calculus of variations'',
{\sl Ann. Inst. Fourier Grenoble} {\bf 23}(1) (1973) 203--267.
(\url{http://eudml.org/doc/74112}).

\bibitem{GGR-2022}
J. Gomis, A. Guerra IV, and N. Rom\'an-Roy,
``Multisymplectic constraint analysis of scalar field theories, Chern--Simons gravity, and bosonic string theory'',
{\sl Nucl. Phys. B} {\bf 987} (2023) 116069.
(\url{https://doi.org/10.1016/j.nuclphysb.2022.116069}).

\bibitem{GIMMSY}
M.J. Gotay, J. Isenberg, J.E. Marsden, and R. Montgomery, ``Momentum maps and classical relativistic fields. I. Covariant theory'', arXiv:physics/9801019 [math-ph] (2004).

\bibitem{GP-2002}
X. Gr\`acia and J.M. Pons,
``Symmetries and infinitesimal symmetries of singular differential equations'',
{\sl J. Phys. A: Math. Gen.} {\bf 35}(24) (2002) 5059.
(\url{https://doi.org/10.1088/0305-4470/35/24/306}).

\bibitem{GR-2023}
A. Guerra IV and N. Rom\'an-Roy,
``More insights into symmetries in multisymplectic field theories'',
{\sl Symmetry} {\bf 2023}, 15, 390 (2023).
(\url{https://doi.org/10.3390/sym15020390}).

\bibitem{HK-04}
F. H\'elein and J. Kouneiher,
``Covariant Hamiltonian formalism for the calculus of variations with several variables: Lepage--Dedecker versus De Donder--Weyl'',
{\sl Adv. Theor. Math. Phys.} {\bf 8} (2004), 565--601.

\bibitem{JP-2015}
R. Jackiw, So-Young Pi,
``Fake Conformal Symmetry in Conformal Cosmological Models'',
{\sl Phys. Rev. D} {\bf 91} (2015) 067501.
(\url{https://doi.org/10.1103/PhysRevD.91.067501}).

\bibitem{Kanatchikov1}
I.V. Kanatchikov,
``Canonical structure of classical field theory in the polymomentum phase space'',
{\sl Rep. Math. Phys.} {\bf 41} (1998) 49-90.
(\url{https://doi.org/10.1016/S0034-4877\%2898\%2980182-1}).

\bibitem{Kanatchikov2}
I.V. Kanatchikov,
``De Donder-Weyl Hamiltonian formulation and precanonical quantization of vielbein gravity'',
{\sl J. Phys.: Conf. Ser.} {\bf 442} (2013) 012041.
(\url{https://doi.org/10.1063/1.4791728
}).

\bibitem{KSM-2011}
I. Kol\'ar, J. Slov\'ak, and P.W. Michor,
{\it Natural Operations in Differential Geometry},
Springer Berlin, Heidelberg, 1993.
(\url{https://doi.org/10.1007/978-3-662-02950-3}).

\bibitem{Krupka}
D. Krupka,
{\it Introduction to Global Variational Geometry}, {\sl Atlantis Studies
in Variational Geometry}, 
Atlantis Press 2015.
(\url{https://link.springer.com/book/10.2991/978-94-6239-073-7}).

\bibitem{Li-2016}
A. Lichnerowicz, {\it Elements of Tensor Calculus},
Dover, New York 2016.
ISBN: 9780486811864.

\bibitem{Mi-2008}
P.W. Michor,
{\it Topics in Differential Geometry},
{\sl Grad. Stud. in Math.} {\bf 93}, Am. Math. Soc., Wien (Austria) 2008. 
ISBN-10: 0821820036.

\bibitem{art:Roman09}
N. Rom\'{a}n-Roy,
``Multisymplectic Lagrangian and Hamiltonian formalisms of classical field theories'',
\textsl{Symm. Integ. Geom. Meth. Appl.}
\textbf{5} (2009) 100. 
 (\url{https://doi.org/10.3842/SIGMA.2009.100}).

 \bibitem{RWZ-2016}
L. Ryvkin, T. Wurzbacher, and M. Zambon,
``Conserved quantities on multisymplectic manifolds'',
{\sl J. Austral Math. Soc.} {\bf 108}(1) (2020) 120--144.
(\url{https://doi.org/10.1017/S1446788718000381}).

\bibitem{Sa-95}
G. Sardanashvily,
{\it Generalized Hamiltonian Formalism for Field Theory. Constraint Systems},
World Scientific Pub. Co. Inc., River Edge, NJ, 1995. (\url{https://doi.org/10.1142/2550}).

\bibitem{book:Saunders89}
D.J. {Saunders}, \emph{The geometry of jet bundles}, London Mathematical
  Society, {\sl Lect. Notes Ser.} {\bf 142}, Cambridge Univ. Press,
  Cambridge, New York, 1989.
(\url{https://doi.org/10.1017/CBO9780511526411}).

\bibitem{Vey-2015}
D. Vey, 
``Multisymplectic formulation of vielbein gravity. De Donder-Weyl formulation, Hamiltonian $(n-1)$-forms'',
{\sl Class. Quantum Grav.} {\bf 32}(9) (2015) 095005.
(\url{https://doi.org/10.1088/0264-9381/32/9/095005}).

\bibitem{We-1935}
H. Weyl, 
``Geodesic fields in the calculus of variation for multiple integrals'',
{\sl Ann. Math.} {\bf 36}(3) (1935) 607--629. 
(\url{https://doi.org/10.2307/1968645}).

}
\end{thebibliography}
\end{document}